\documentclass[conference]{IEEEtran}
\IEEEoverridecommandlockouts
\usepackage{cite}
\usepackage{amsmath,amssymb,amsfonts}
\usepackage{mathrsfs}
\usepackage{algorithmic}
\usepackage{graphicx}
\usepackage{textcomp}
\usepackage{enumitem}
\usepackage{subfigure}
\usepackage{epstopdf}
\usepackage{balance}
\usepackage{algorithm}
\usepackage{bm}
\usepackage{color,soul}
\usepackage{siunitx} 
\usepackage{cancel}
\usepackage{multicol}
\usepackage{verbatim}
\usepackage{booktabs}
\usepackage{siunitx}
\usepackage{graphics}
\usepackage{multirow}
\usepackage{array}
\usepackage{color}
\usepackage{colortbl}
\usepackage{url}
\usepackage{endnotes}
\usepackage{microtype}

\def\BibTeX{{\rm B\kern-.05em{\sc i\kern-.025em b}\kern-.08em
		T\kern-.1667em\lower.7ex\hbox{E}\kern-.125emX}}

\definecolor{mygray}{gray}{.9}
\definecolor{mypink}{rgb}{.99,.91,.95}
\definecolor{mycyan}{cmyk}{.3,0,0,0}

\begin{document}

\title{Reliable Wide-Area Backscatter via Channel Polarization\\}

%
%

\author{\IEEEauthorblockN{Guochao Song,  Hang Yang, Wei Wang, Tao Jiang}
	\ Huazhong University of Science and Technology, China
	 \\ Email: \{sgc, hangyang, weiwangw, taojiang\}@hust.edu.cn
 }

\maketitle
\begin{abstract}	
A long-standing vision of backscatter communications is to provide long-range connectivity and high-speed transmissions for batteryless Internet-of-Things (IoT). Recent years have seen major innovations in designing backscatters toward this goal. Yet, they either operate at a very short range, or experience extremely low throughput. This paper takes one step further toward breaking this stalemate, by presenting PolarScatter that exploits channel polarization in long-range backscatter links. We transform backscatter channels into nearly noiseless virtual channels through channel polarization, and convey bits with extremely low error probability. Specifically, we propose a new polar code scheme that automatically adapts itself to different channel quality, and design a low-cost encoder to accommodate polar codes on resource-constrained backscatter tags. We build a prototype PCB tag and test it in various outdoor and indoor environments. Our experiments show that our prototype achieves up to 10$\times$ throughput gain, or extends the range limit by 1.8$\times$ compared with the state-of-the-art long-range backscatter solution. We also simulate an IC design in TSMC 65 nm LP CMOS process. Compared with traditional encoders, our encoder reduces storage overhead by three orders of magnitude, and lowers the power consumption to tens of microwatts.

\end{abstract}


\section{Introduction} 

Backscatter has recently become a prominent contender in Internet-of-Things (IoT) due to the extremely low energy consumption that can work in a batteryless manner as well as its extremely low cost at a few cents~\cite{Ambientbackscatter,Ekhonet}. It has always been dreamed in developing backscatter technologies to deliver relatively high-rate, reliable transmissions with wide-area coverage. These features would promise its applications in battery-free surveillance~\cite{videoscatter}, precision agriculture~\cite{xu2018practical}, ground water quality measurement and inventory management in large warehouses~\cite{NetScatter}.

Despite growing attempts and extensive efforts, current backscatter systems either operate at a short range~\cite{Backfi,Ekhonet}, or provide very low throughput~\cite{NetScatter,talla2017lora,peng2018plora}. 
The culprit lies in the fundamental backscatter paradigm that backscatter tags do not emit signals but only scatter ambient signals, making backscatter links suffer from extremely poor channel quality when communicating over long distances~\cite{xu2018practical}.

Through careful measurements, we observe that there exists a disparity in long-range backscatter transmissions: a large portion of lost packets contain only a mismatched small number of error bits. It indicates that current designs do not take full advantage of the backscatter channel, making it largely underutilized. As a symbol in long-range backscatter usually lasts for milliseconds, and the channel may vary during the transmission of a single backscatter packet. For a long-range backscatter link with extremely poor channel quality, a subtle variance in the channel may cause catastrophic impact on packet decoding. 

This inspires us to exploit channel polarization to divide the long lasting backscatter channel into multiple subchannels. Particularly, we transform the physical backscatter channel into a set of virtual outer channels, which are polarized to be either extremely reliable or very unreliable. The reliable ones are used to convey information bits. To this end, we propose PolarScatter that enables reliable wide-area backscatter beyond the current limit by exploiting channel polarization based on polar codes.



Existing polar codes, however, are primarily designed for high-end devices with active radios. They consume significant computational resources and power, and need accurate channel estimation. Accommodating polar codes to power-constrained and low-complexity backscatter tags entails several practical challenges.
The first challenge arises from ineffective code rate adaptation in long-range backscatter. Traditional code rate adaptation requires the receiver to program each client to operate on a suitable code rate based on its received signal quality. However, in fact, it is particularly hard to obtain the received signal quality since the backscatter signal is drowned by noise.   We address this challenge by introducing \textit{Sozu}\footnote[1]{Sozu is a Japanese fountain device aotumatically pivoting to one side of its balance point.}
polar codes. In this scheme, backscatter tags initially encode information bits with a high code rate, and then add redundant bits in an extra transmission according to the decoding requirement of the receiver which can be estimated based on frozen bits error ratio (FBER).  
With \textit{Sozu} polar codes, tags can automatically adjust to a suitable effective code rate for different channel quality. 

Second, conventional polar codes incur excessive encoding overhead for resource-constrained backscatter tags. The problem lies in that the storage of the coding matrix leads to overwhelming overhead. We tackle this challenge by leveraging the fact that the coding matrix can be generated by matrix iterations. Instead of storing all coding matrix for different code lengths, we simply save the coding matrix with the minimum length, and compute the coding matrix of long codes by iterations. Furthermore, to reduce the computational complexity, we push the permutation operation of encoding into the receiver, thereby leaving tags as simple as possible.

Finally, to obtain excellent decoding performance, the decoder requires accurate calculation of log-likelihood ratio (LLR) of received bits. Conventional methods compute LLR depending on the power of signal and noise. However, it is difficult to estimate the signal power in long-range backscatter, as the backscatter signal is drowned by noise. Thus, we need to devise a new metric to calculate the LLR  without the signal power.  Note that current long-range backscatters leverage LoRa signals as the excitation signals. To decode the LoRa signal, the receiver performs an fast Fourier transform (FFT) on the multiplication of the incoming chirp and a down chirp, which leads to a peak in an FFT frequency bin. Instead of directly computing the probability of the peak appearance in a certain FFT bin, we compute the probability of all other bins without peaks, which merely depends on the noise power. In this way, we can accurately estimate LLR without the signal power.


To verify our design, we build  a hardware prototype of PolarScatter using an FPGA platform and a customized backscatter analog frontend board.  We conduct extensive experiments to test the performance of  PolarScatter in line-of-sight (LoS) and non-line-of-sight (NLoS) scenarios.  Our experiment results show that PolarScatter achieves up to 10$\times$ throughput gain, or extends the range limit by 1.8$\times$ compared with the state-of-the-art long-range backscatter. We also design an encoder integrated circuit (IC) based on TSMC 65 nm LP CMOS process, and evaluate resource cost and power consumption using industry-standard EDA tools. Our results show that compared with the traditional encoder, our encoder consumes merely $0.2\%$ storage overhead and lowers the power consumption from 1.78~mW to 71 $\mu$W.

We summarize our contributions as follows:  
\begin{itemize}
	\item  We bring in channel polarization to provide long-range connectivity and high-speed transmissions for batteryless backscatter. We design \textit{Sozu} polar codes to best exploit the link capacity and automatically adjust to a suitable effective bit rate for different channel quality.  
	\item  We propose a low-cost polar encoder to accommodate polar codes on resource-constrained backscatter tags.
	\item  We devise a new metric to calculate LLR. We can leverage this metric to accurately perform polar decoding.
	\item  We prototype our design, and the experimental results demonstrate our merits in high throughput for long-range transmissions, low encoding overhead and power consumption.
\end{itemize}

The rest of the paper is organized as follows. We begin with our motivation in Section~\ref{sec:motivation}. We elaborate on system design in
Section~~\ref{sec:design}, followed by our implementation in Section~\ref{sec:Implementation}. We show the evaluation results in Section~\ref{sec:Evaluation}. Related work is reviewed in Section~\ref{sec:relatedwork}. Finally, we conclude the paper in Section~\ref{sec:Conclusion}.

\section{Exploiting Channel Polarization in Long-range Backscatter} \label{sec:motivation}
\begin{figure}[t]
	\begin{minipage}{0\linewidth}
		\centering
		\subfigure[Packet level]{	
			\label{PRR}	
			\includegraphics[width=1.85in]{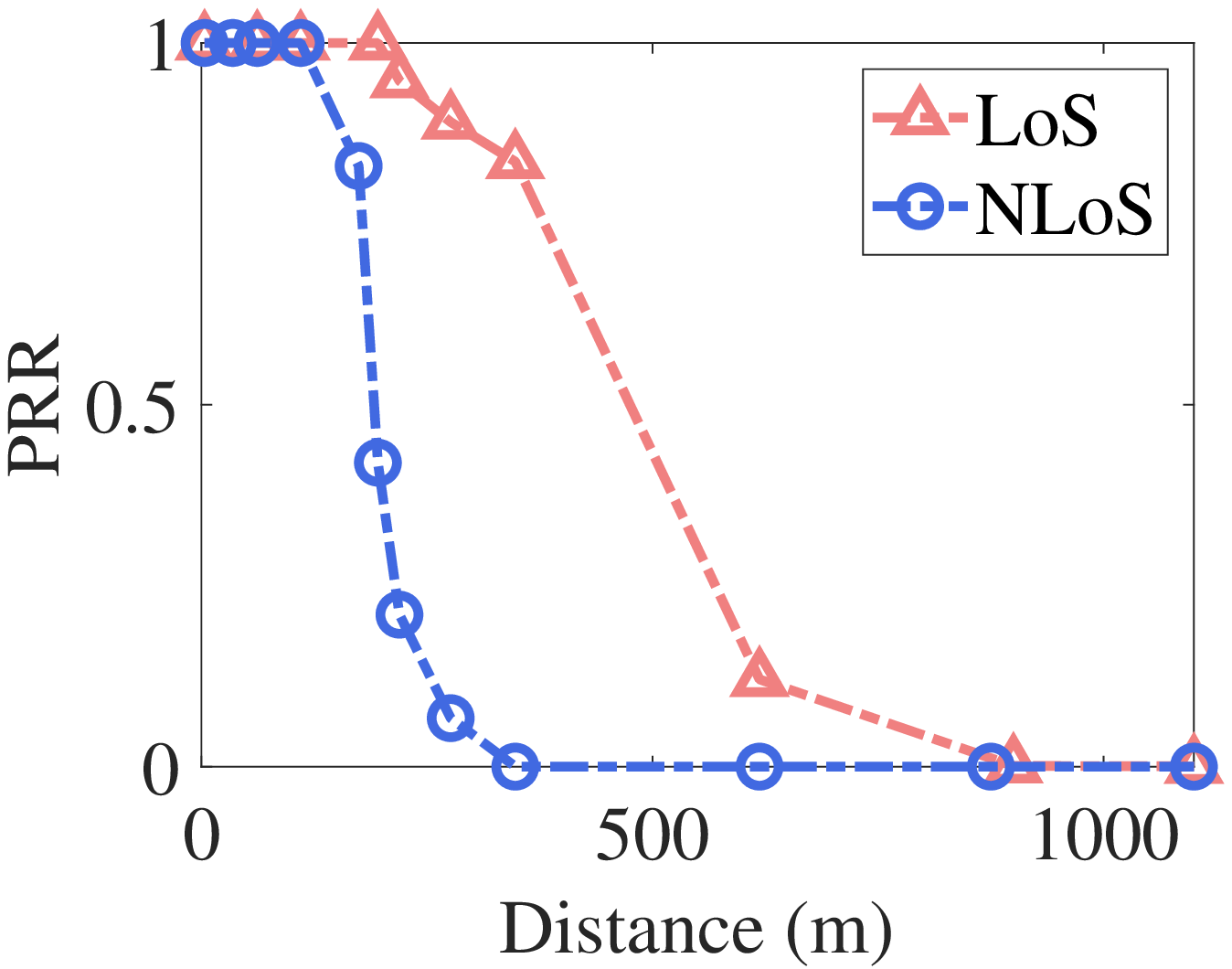}}
	\end{minipage}%
	\begin{minipage}{2\linewidth}
		\hspace{-1.75in} 
		\centering
		\subfigure[Byte level]{	
			\label{BRR}	
			\includegraphics[width=1.85in]{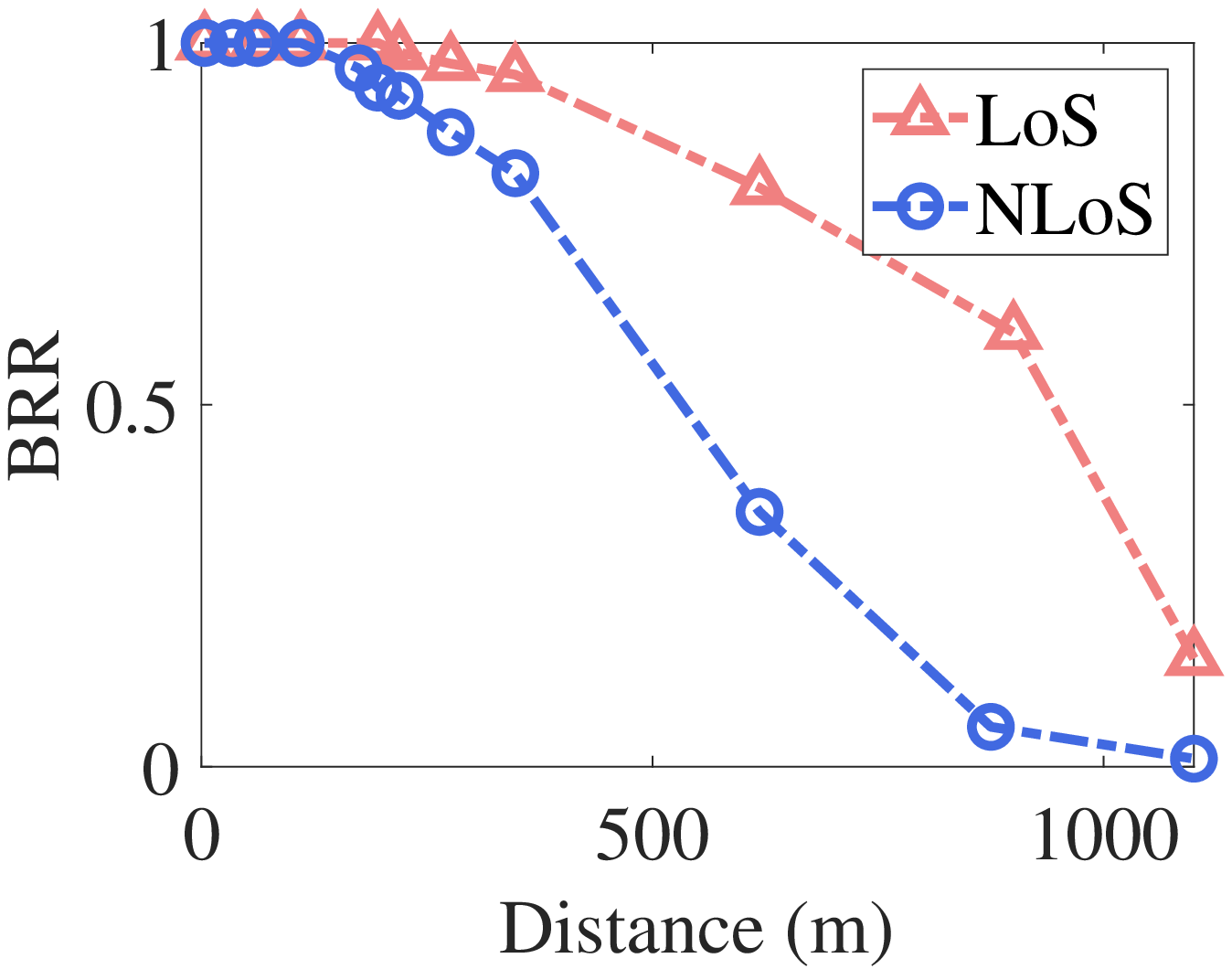}}
	\end{minipage}%
	\caption{Decoding performance under different communication distances.} 
	\label{PRRBRR}
	\vspace{-0.5cm}
\end{figure}

\textbf{Poor decoding performance in long-range backscatter.}
Some long-range backscatter technologies, like LoRa Backscatter\cite{talla2017lora} and PLoRa\cite{peng2018plora}, have been developed to support  kilometer-level communication distance. They leverage narrow chirp spread spectrum and high sensitivity receiver to decode weak backscatter signals below the noise floor. However, the decoding performance is vulnerable to significant path loss and multi-path effect from surrounding obstacles, resulting in extremely low throughput.

To better understand this, we conduct a series of measurements using PLoRa backscatter\cite{peng2018plora} in LoS and NLoS scenarios, as shown in Fig.~\ref{MAP}. For each experiment, we let an active LoRa node continuously transmits packets with the maximum output power (20 dBm), and a PLoRa tag\cite{peng2018plora} backscatters the LoRa packets. The packet size is set with an 8-Byte payload and a 2-Byte cyclic
redundancy check (CRC). The PLoRa tag is placed near the active LoRa node with 1 m distance. We change the transmission distance by moving a gateway. We report the packet reception rate (PRR)  and byte reception rate (BRR).

Fig.~\ref{PRR} shows the measured average PRR  at different communication distances. We see that the PRR is extremely low when the transmission distance is more than 500 m in the LoS scenario. Even worse, the PRR further decreases in the NLoS scenario. The measurement results reveal the backscatter link capacity is largely underutilized due to very low PRR.
We further investigate the BRR in Fig.~\ref{BRR}. It can be seen there exists a large number of correct bytes in long communication distance. 
The reason is that a symbol in long-range backscatter usually lasts for milliseconds, and the channel may vary during the transmission of a single backscatter packet. Even worse, the long-range backscatter channels normally suffer from severe fading, and thus a subtle channel variance may lead to catastrophic impact on packet decoding.
The observations motivate us to exploit channel polarization to make effective utilization of correct bytes, thereby providing reliable and relatively high-rate transmissions across large areas.

\begin{figure}[t]
	\centering
	\includegraphics[width=3.0in]{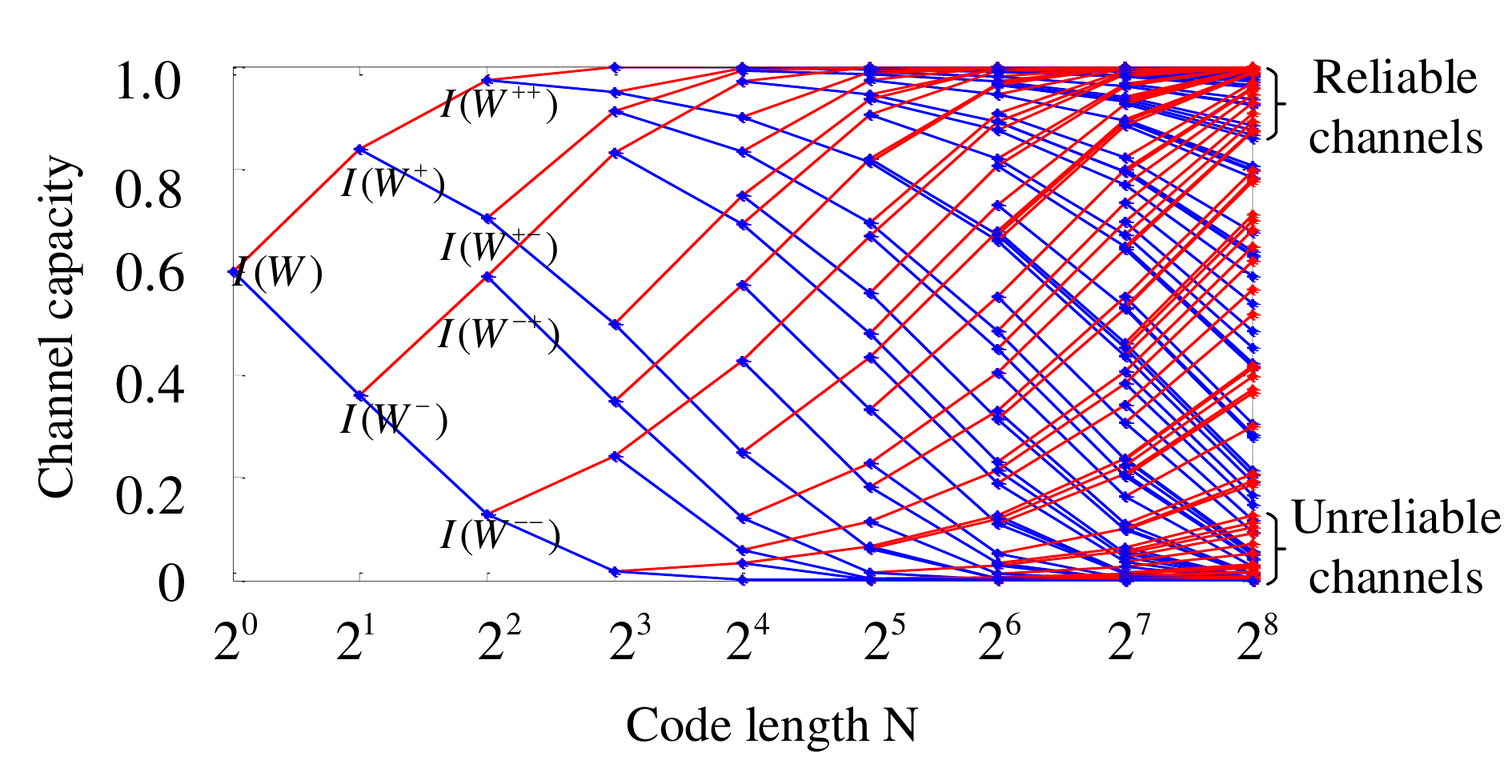}  \vspace{-0.3cm}
	\caption{Evolution of channel polarization.}
	\label{polarcode} 
	\vspace{-0.3cm}
\end{figure}

\textbf{Channel polarization.} 
Channel polarization transforms $N$ copies of physical channels to a set of $N$ virtual channels. As the $N$ becomes larger, the virtual channels tend to either have high reliability or low reliability (\textit{i.e.}, they polarize). Fig.~\ref{polarcode} shows the evolutionary trajectories of channel polarization from $N=2^0$ to $N=2^8$, where each physical channel is assumed to be a binary erasure channel (BEC) with the channel capacity $0.6$. We refer high-reliability or low-reliability channels as ``reliable'' and ``unreliable'' channels which are marked with red and blue lines, respectively. The unreliable virtuals are assigned with the fixed bits (called frozen bits). In contrast, information bits (called free bits) are transmitted over the reliable channels with extremely low error probability.  
Thus, by leveraging reliable channels, we can pick out the correct bytes to make effective utilization of backscatter link capacity.

\section{PolarScatter Design}  \label{sec:design}

In this section, we first describe an overview of PolarScatter. Then, we propose \textit{Sozu} polar code, which can automatically adjust to a suitable code rate for different channel quality. Next, we present a low-cost encoder, and discuss how to accurately decode. Finally, we outline the PolarScatter's protocol design.

\subsection{Design Overview}
\begin{figure}[t]
	\centering
	\includegraphics[width=3.5in]{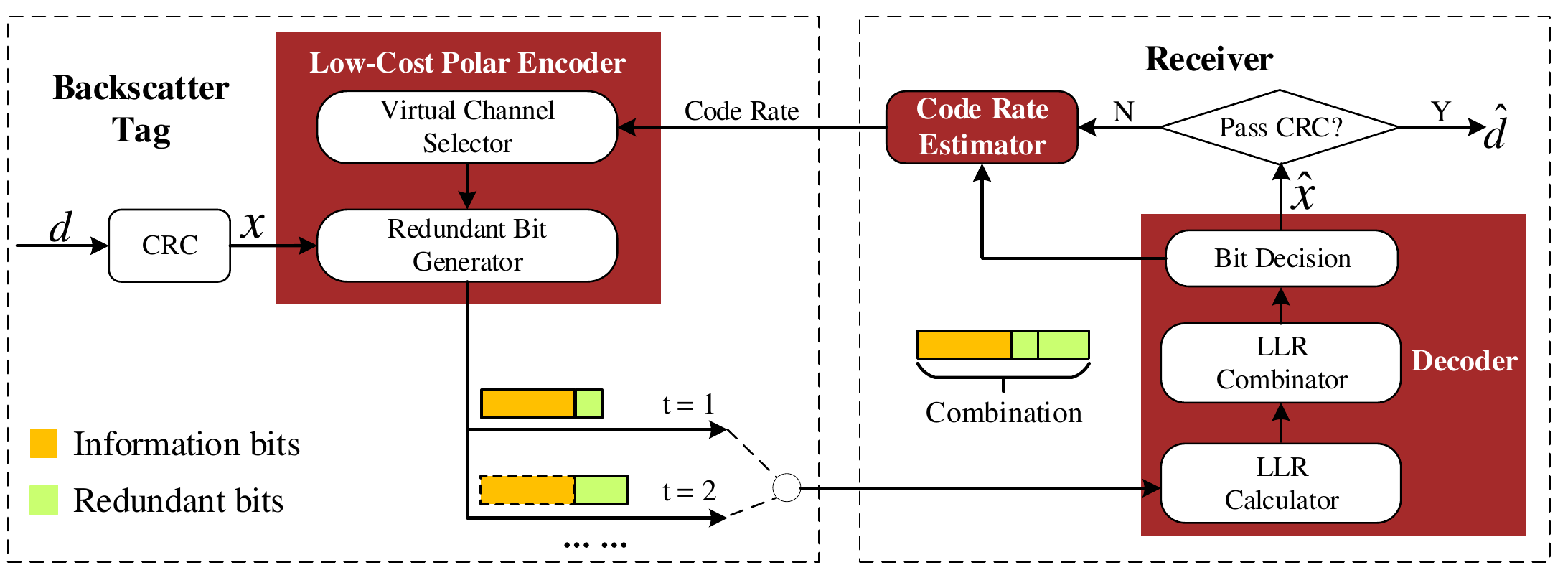} 
	\caption{Architecture of PolarScatter design.}
	\label{architecture}\vspace{-0.4cm}
\end{figure}

PolarScatter is a  cross-layer design that leverages channel polarization to enable high throughput in long-range backscatter.
The crux of PolarScatter is a channel adaptive coding scheme that maps the tag's information bits into reliable virtual channels and automatically adjusts to a suitable code rate for different channel quality. 
Fig.~\ref{architecture} shows the system flow of PolarScatter transmission and reception. PolarScatter extends existing backscatter by adding the following three components.

\begin{itemize}
	\item  \textbf{Low-cost Polar Encoder} residing atop the tag's physical layer. 
	PolarScatter adds the polar encoder at the tag's physical layer without modifying inherent blocks such as CRC and modulation modules. The encoder includes two modules: a virtual channel selector and a redundant bit generator. The virtual channel selector picks reliable channels out of all virtual channels to transmit information bits. 
	The redundant bit generator computes the redundant bits based on the reliable virtual channels.
	\item  \textbf{Polar Decoder}  incorporated in the decoding pipeline. 
	The polar decoder works at the physical layer of the receiver and is composed of three modules: an LLR calculator, an LLR combinator and a bit decision module. The goal of the LLR calculator is to estimate the LLR of information bits for each packet. The LLR combinator merges the LLR of information bits of all packet. The bit decision determines the information bits based on the combined LLR.   
	\item  \textbf{Code Rate Estimator}  operating on the link layer of the gateway. 
	PolarScatter pushes the code rate estimator into gateway to reduce the burden of  tags.  The  code rate estimator aims to obtain a proper code rate that can meet the requirement of decoding, which is estimated based on FBER.
\end{itemize}

\subsection{Channel Adaptive Coding Design}

The core and challenging part of PolarScatter is to design a polar code scheme that can synthesize $K$ reliable virtual channels out of $N$ physical channels for different channel quality. The $K$ reliable virtual channels are used to convey information bits. Traditional code rate adaptation approaches depend on received signal quality. Since the received signal is usually drowned by noise in long-range backscatter, it is difficult to estimate the received signal quality. Thus, these approaches are not suitable for long-range backscatter.

To overcome this challenge, we propose \textit{Sozu} polar codes. The key idea of \textit{Sozu} polar codes is to encode information bits in systematic codes, and adds redundant bits according to the requirements of decoding. In this way, PolarScatter can automatically reach the optimal code rate corresponding to the channel quality.

In particular, the \textit{Sozu} coding scheme includes two stages.
In the first stage, the tag encodes the informations bits in a systematic code with 3/4 code rate, and then transmits information bits and redundant bits to the receiver. After receiving the first frame, the gateway performs polar decoding. If information bits are successfully decoded, the receiver informs the tag to terminate its transmission. Otherwise, the receiver estimates a code rate that can meet the requirement of decoding based on FBER, and feeds back the code rate to the tag. 
In the second stage, the tag encodes information bits with the requested code rate, and 
then transmits redundant bits to the receiver. After receiving the second frame, the receiver combines all redundant bits from different frames to perform decoding. With redundant bit combination, the decoding performance is significantly boosted.
Next, we introduce the code rate estimation.

\textbf{ FBER-aware Code Rate Estimation.}
We leverage FBER to estimate the code rate.
Recall that the unreliable virtual channels are assigned with frozen bits,  which are usually set as the bit ``0''. The frozen bits are known to both the receiver and the transmitter, and thus we can treat the frozen bits as prior knowledge to measure channel quality. When the channel quality is poor, the frozen bits may be incorrectly decoded. The larger the FBER is, the worse the channel quality is. To accurately characterize the mapping from FBER to code rate, we turn to an empirical study. In the experiment, we test the minimum code rate that makes information bits be correctly decoded at different FBER. The test results are shown in Fig.~\ref{LLRvsCodelength}. We can see that the relationships between FBER and code rate are the same for different code lengths. Besides, at the same FBER, the code rate fluctuates slightly. 
Thus, we can establish a discrete mapping from FBER to code rate in the receiver.
We set four code rates (2/3, 1/2, 1/4, and 1/8), and the corresponding FBER ranges are $0.1-0.3$, $ 0.3-0.5$,  $0.5-0.7$, and $0.7-1$, respectively.    


\subsection{Low-cost Encoder} 
So far, we have introduced \textit{Sozu} polar codes. We now discuss how to accommodate the corresponding encoder on resource-constrained backscatter tags. 

\textbf{Selecting reliable virtual channels.}
The first step of encoder is to map $K$ information bits into reliable virtual channels. 
Traditionally, the reliability of each virtual channel can be estimated by a basic butterfly operation\cite{arikan2009channel,SCL}. 
Unfortunately, the butterfly operation entails extremely high computational overhead, making it unaffordable to resource-constraint backscatter tags. 

We address this challenge based on a key observation that while the reliability order of virtual channels is distinct at different code rates, the indices of $K$ most reliable channels at a low code rate are the same as that at a higher code rate.
To illustrate it, we present an example of the channel reliability order at different code rates with a code length of 32 in Table~\ref{channelindex}, where the indices of reliable channels are highlighted.  We can see the indices of the most reliable channels at code rates of 1/4 and 1/8 are the same as those at a higher code rate of 1/2. Thus, we can only save the reliability order of virtual channels for the highest code rate (\textit{i.e.}, 3/4 code rate), and select the top $K$ indices as the set of the reliable channels for other code rates. 

\begin{figure}
	\begin{minipage}[t]{0.23\textwidth}
		\centering
		\includegraphics[width=1.9in]{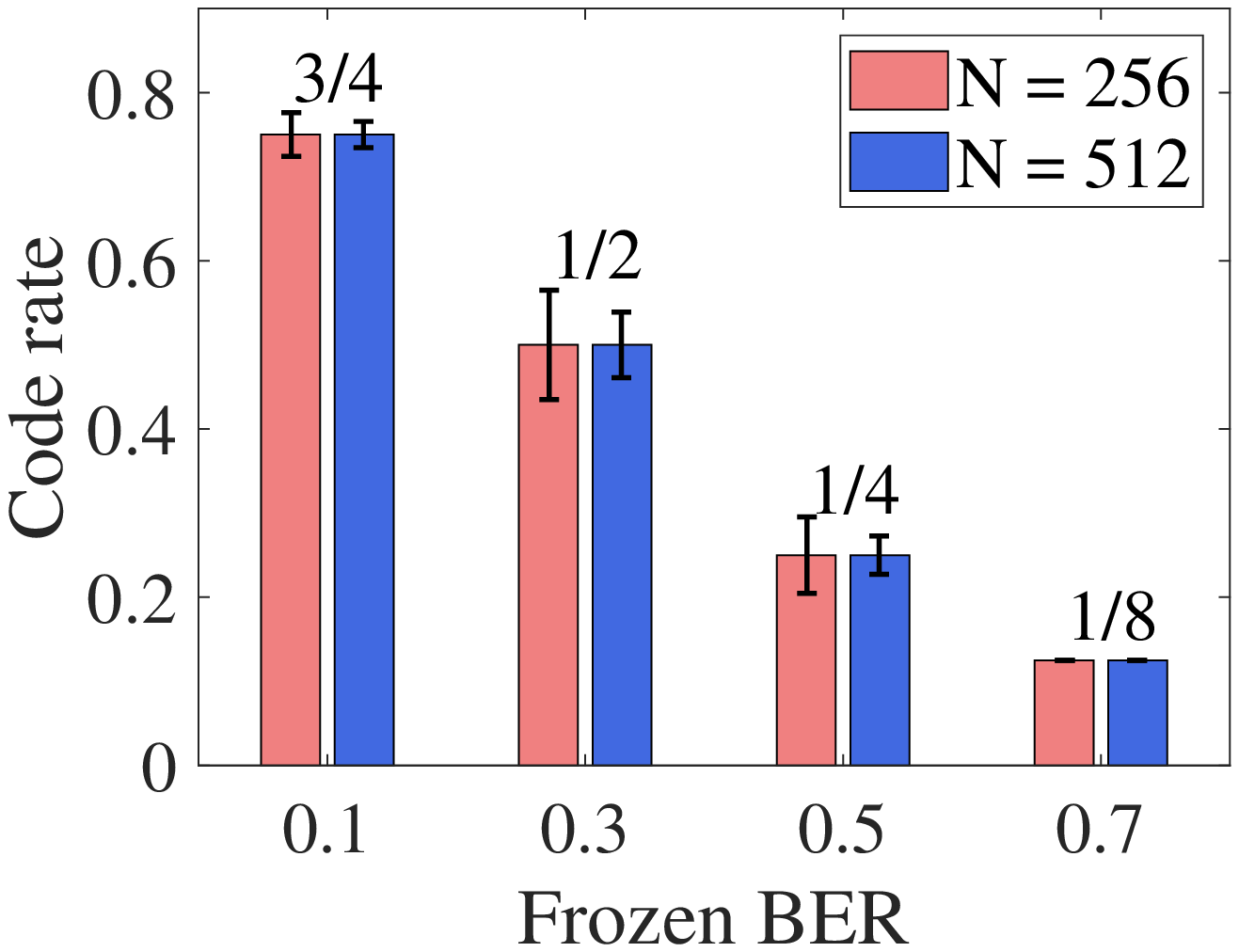}\vspace{-0.1cm}
		\caption{Relationship between code rate and frozen BER.}
		\label{LLRvsCodelength}
	\end{minipage}
	 \vspace{-0.2cm}
	\begin{minipage}[t]{0.23\textwidth}
		\centering
		\includegraphics[width=1.4in]{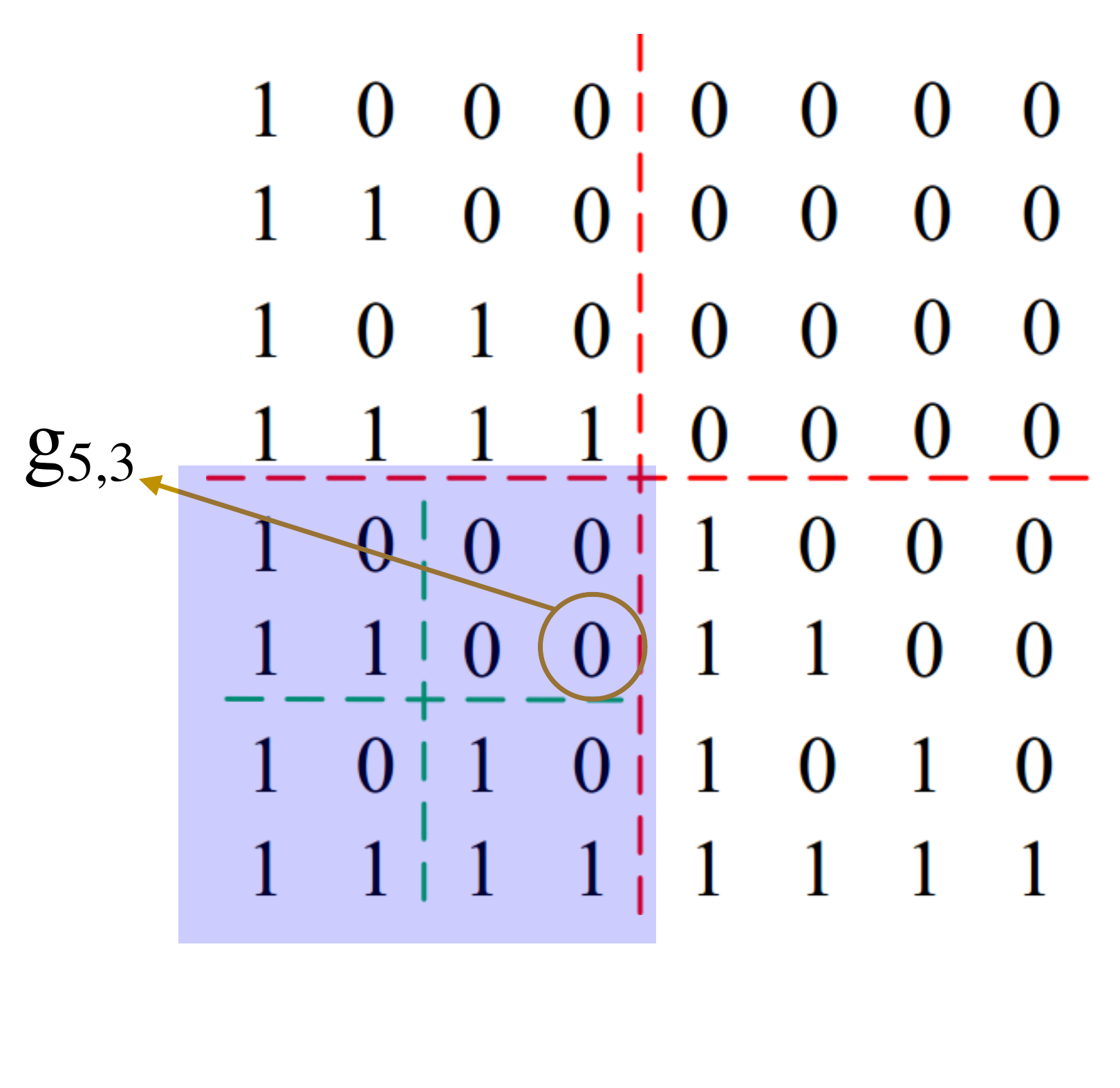}\vspace{-0.1cm}
		\caption{Coding matrix $G_8$.}
		\label{Codingmatrix}
	\end{minipage}
	\vspace{-0.0cm}
\end{figure}

\textbf{Generating redundant bits.} Now that we know the set of reliable channels, let us focus on how to compute the redundant bits. 
Let $\mathcal{A}$ and $\mathcal{B}$ denote the sets of $K$ reliable channel indices and $N-K$ unreliable virtual indices, respectively.
According to \cite{systematicpolar,SPCLetters}, the redundant bits $\mathbf{x}_r$  can be computed by
\begin{equation}\label{encoder3}
\mathbf{x}_r =  \mathbf{G}_\mathcal{AB} \mathbf{G}_\mathcal{A} \mathbf{x},
\end{equation}
where $\mathbf{x}$ denotes information bit vector, and $\mathbf{G}_\mathcal{AB} = (g_{i,j}) \in \mathbb{R}^{(N-K) \times K}$ and $\mathbf{G}_\mathcal{A} = (g_{k,l}) \in \mathbb{R}^{K \times K}$ are the submatrices of the coding matrix $\mathbf{G}_N$, respectively, with $i, l, k \in \mathcal{A}$ and $ j\in \mathcal{B}$.
The coding matrix $\mathbf{G}_N$ is denoted by
\begin{equation}\label{Gn}
\mathbf{G}_N = \mathbf{F}^{\otimes n} = \left[\begin{matrix} 1 & 0 \\ 1 & 1 \end{matrix}\right]^{\otimes n},
\end{equation} 
where $\mathbf{F}$ is the kernel matrix and $\mathbf{F}^{\otimes n}$ is the $n$th Kronecker  power of $\mathbf{F}$.

According to the above equation, we need to know the coding matrix $\mathbf{G}_N$ to compute the redundant bits. Conventional polar encoders store the entire coding matrix $\mathbf{G}_N$ to accelerate encoding.
However, the storage of the coding matrix $\mathbf{G}_N$ may consume excessive memory, which is very challenging for resource-constrained backscatter tags.  For example, when the code length is 512, it requires 256 KB memory to store the coding matrix.  

Instead of storing the entire code matrix, we iteratively generate each element by taking advantage of the special structure of this matrix.
We introduce how to compute the element, $g_{m,n} \triangleq [\mathbf{G}_N]_{m,n}$, by giving an example, where $N = 8$, and the element we want to obtain is $g_{5,3}$. 
As shown in  Fig.~\ref{Codingmatrix}, we divide the matrix into four parts, where each part can be represented with two binary bits: ``$00$'' corresponds to the upper left part, ``$01$''  the upper  right part, ``$10$''  the lower left part, and ``$11$''  the lower right part, respectively. 
We denote the row and column indices of $g_{5,3}$ by binary representations, ``$1 0 1$'' and ``$0 1 1$'', respectively. In the first iteration, the first bit of the $5$ and $3$ are ``$1$'' and ``$0$'', so the lower left part, \textit{i.e.}, the blue part, is the selected submatrix. In the second iteration, we further divide the selected submatrix into four parts. In this iteration, as the two bits are ``$0$'' and ``$1$'', the upper right part of the blue submatrix is selected. In the third iteration, the two bits are ``$1$'' and ``$1$'', so the lower right part is selected, which is the circled number. We observe that the upper right part of every submatrix in each iteration are all zeros. It indicates that we can quickly obtain the value of $g_{m,n}$ by the following rule: when there exits an index $i$, where the $i$-th bit in $m$'s binary representation is ``$0$'' and the $i$-th bit in $n$'s binary representation is ``$1$'', $g_{m,n}=0$; otherwise, $g_{m,n}=1$. Since the above computations are binary operations, the computational complexity is extremely low.

\begin{table}
	\renewcommand\arraystretch{1.2}
	\sethlcolor{mypink} 
	\centering
	\caption{Channel indices in a Descending Order of Reliability.}
	\begin{tabular}{|c|c|}
		\hline
		\multicolumn{1}{|c|}{$CR$}& \multicolumn{1}{c|}{$indices$}\\
		\hline
		\multirow{2}*{$1/2$}&\hl{$32,31,30,28,24,16,29,27,26,23,22,15,20,14,12,25$} \\
		&$8,21,19,13,18,11,10,7,6,17,4,9,5,3,2,1$\\
		\hline
		\multirow{2}*{$1/4$}&\hl{$32,31,30,28,24,16,29,27,$}{$26,23,22,15,20,14,25,12$}\\
		&$8,21,19,13,18,11,10,7,6,4,17,9,5,3,2,1$\\
		\hline
		\multirow{2}*{$1/8$}&\hl{$32,31,30,28,$}{$24,29,16,27,26,23,22,15,20,14,12,25$}\\
		&$8,21,19,18,13,11,10,7,6,4,17,9,5,3,2,1$ \\
		\hline
	\end{tabular}
	\label{channelindex}
	\vspace{-0.4cm}
\end{table}

In this way, we can obtain each element of $\mathbf{G}_\mathcal{A}$ and $\mathbf{G}_\mathcal{AB}$. Next, we perform the matrix product $\mathbf{G}_\mathcal{AB} \mathbf{G}_\mathcal{A} \mathbf{x}$. In order to save memory space, we first calculate a temporary vector, $\mathbf{x}_\mathcal{A} = \mathbf{G}_\mathcal{A} \mathbf{x}$.
We allocate $K$-bit memory to save the first row of $\mathbf{G}_\mathcal{A}$ and calculate its product with $\mathbf{x}$, which is the first element of $\mathbf{x}_\mathcal{A}$.
Then, we update the $K$-bit memory with the second row of $\mathbf{G}_\mathcal{A}$ and compute the second element of $\mathbf{x}_\mathcal{A}$. 
The above steps are repeated until all elements in $\mathbf{x}_\mathcal{A}$ are obtained.
Finally, we perform a similar process to compute $\mathbf{x}_r = \mathbf{G}_\mathcal{AB} \mathbf{x}_\mathcal{A}$.

\subsection{\textit{Sozu} Polar Decoder}

We now describe how to perform \textit{Sozu} polar decoding based on received backscatter signals.
To enable reliable backscatter link transmission, the gateway needs to timely feed back a requested code rate to the tag after first decoding. 
We choose the belief propagation (BP)  algorithm to perform decoding, since it achieves excellent decoding performance and maintains advantages in parallelism and low latency\cite{BP}.
Our \textit{Sozu} polar decoder extends the legacy BP algorithm by adding two key components: LLR calculation and LLR combination. Next, we detail \textit{Sozu} decoder.

\begin{figure}[t]
	\centering
	\includegraphics[width=3.7in]{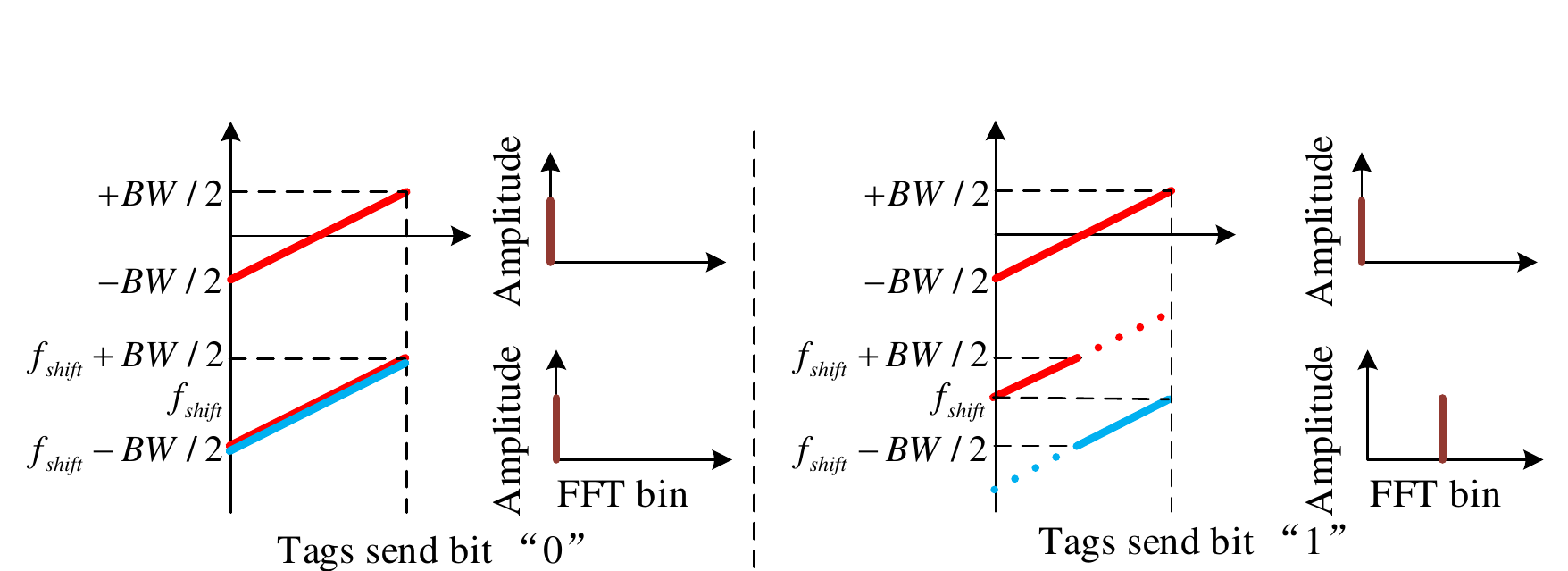} \vspace{-0.1cm}
	\caption{The PLoRa chirp signal ``0'' and chirp signal ``1''. PLoRa tags use two antennas to synthesize a legal LoRa signal. }
	\label{PLoRa}
	\vspace{-0.4cm}
\end{figure}

\textbf{LLR calculation.}
The first step of \textit{Sozu} decoding is to compute the LLR of received bits, which is determined by the demodulated signals.
We introduce how to calculate the LLR by taking PLoRa as an example, which can be extended to other types of backscatter. Recall that a PLoRa tag conveys information by modulating the active LoRa chirp into legacy LoRa signal.  The basic idea is to shift the frequency of the incoming active LoRa chirp by two antennas, as shown in Fig.~\ref{PLoRa}. When the tag sends the bit ``0'', the two antennas  shift the incoming LoRa symbol from the center frequency at $f_0$ to another the center frequency  at shift $f_0 + f_{shift}$. Conversely, if the tag transmits the bit ``1'', two antennas shift the center frequency at $f_0$ to the center frequency at shift $f_0 + f_{shift}+BW/2$ and  the center frequency at shift $f_0 + f_{shift}-BW/2$, respectively. 

To demodulate the tag's data, the gateway performs an FFT on the multiplication of the incoming chirp and a down chirp, and an FFT on the multiplication of the tag's chirp and a down chirp.
Suppose the peak location of active node's FFT for $i$ symbol is $s_i$. Next, we leverage $s_i$ as a prior to compute the LLR of the received bits for the tag.     

Mathematically, the value of the $j$th FFT bin for the $i$th symbol from the tag can be defined as
\begin{equation}\label{LLR2}
\hat{{f}}_{i,j} \triangleq \left\{
\begin{array}{lr}
{\sqrt P}  +  {n_j}, if \;  j=s'_i \\
{n_j}, \text{otherwise},\\
\end{array}\right.
\end{equation} 
where $P$ denotes the received signal power, and $n_j$ is the complex Gaussian noise with variance $2\sigma^2$. The $s'_i$ is the peak location of the FFT bin, which can be represented as
\begin{equation}\label{FFTpeak}
s'_i = \left\{
\begin{array}{lr}
s_i, if \; \text{the tag transmits bit ``0''}\\
\overline{s}_i = (s_i+N)  \bmod {N/2}, if \;\text{the tag transmits bit ``1''}. 
\end{array}\right.
\end{equation} 
For brevity, we denote the FFT bin vector of the $i$th received symbol as $\hat{\mathbf{f}}_i$.

We denote the event that the peak appears in ${j}$th bin of FFT  as $A_{j}$, and denote the event that the peak does not occur in ${j}$th bin of FFT as $B_{j}$. To compute the probability $p(A_j)$, we need to know the $P$. As mentioned earlier, it is difficult to obtain the $P$, since the backscatter signal is drowned by noise.

We observe that the event $A_{s_i}$ is equivalent to the event  $B_{\overline{s}_i}$,  and the event $A_{\overline{s}_i}$ is  equivalent to the event $B_{s_i}$. Instead of estimating the probabilities of $A_{s_i}$  and $A_{\overline{s}_i}$, we can instead compute the probabilities of  $B_{\overline{s}_i}$ and $B_{s_i}$ without knowing $P$. Note that if a variable obeys a complex Gaussian distribution, its magnitude follows the Rayleigh distribution\cite{rayleigh}.
Thus, given $\hat{\mathbf{f}}_i$ and $s_i$, the condition probability pair can be written as
\begin{equation}\label{LLR3}
\begin{aligned}
p(\hat{\mathbf{f}}_i,s_i|x_i=0) = p(B_{\overline{s}_i}) =  \dfrac{|\hat{{f}}_{i,\overline{s}_i}|}{\sigma^2}  \exp(-\dfrac{|\hat{{f}}_{i,\overline{s}_i}|^2}{2\sigma^2}),\\
p(\hat{\mathbf{f}}_i,s_i|x_i=1) = p(B_{s_i}) =  \dfrac{|\hat{{f}}_{i,s_i}|}{\sigma^2} \exp(-\dfrac{|\hat{{f}}_{i,s_i}|^2}{2\sigma^2}).\\ 
\end{aligned}
\end{equation}

Based on (\ref{LLR3}),  the LLR function of the received bits for the $i$th received bit can be denoted as
\begin{equation}\label{LLR1}
\begin{split}
L_{i} = \ln \dfrac{p(\hat{\mathbf{f}}_i,s_i|x_i=0)}{p(\hat{\mathbf{f}}_i,s_i|x_i=1)} = \dfrac{|\hat{{f}}_{i,s_i}|^2-|\hat{{f}}_{i,\overline{s}_i}|^2}{2\sigma^2} \ln \dfrac{|\hat{{f}}_{i,\overline{s}_i}|}{|\hat{{f}}_{i,s_i}|}.
\end{split}
\end{equation}

\begin{figure}[t]
	\centering
	\includegraphics[width=3.4in]{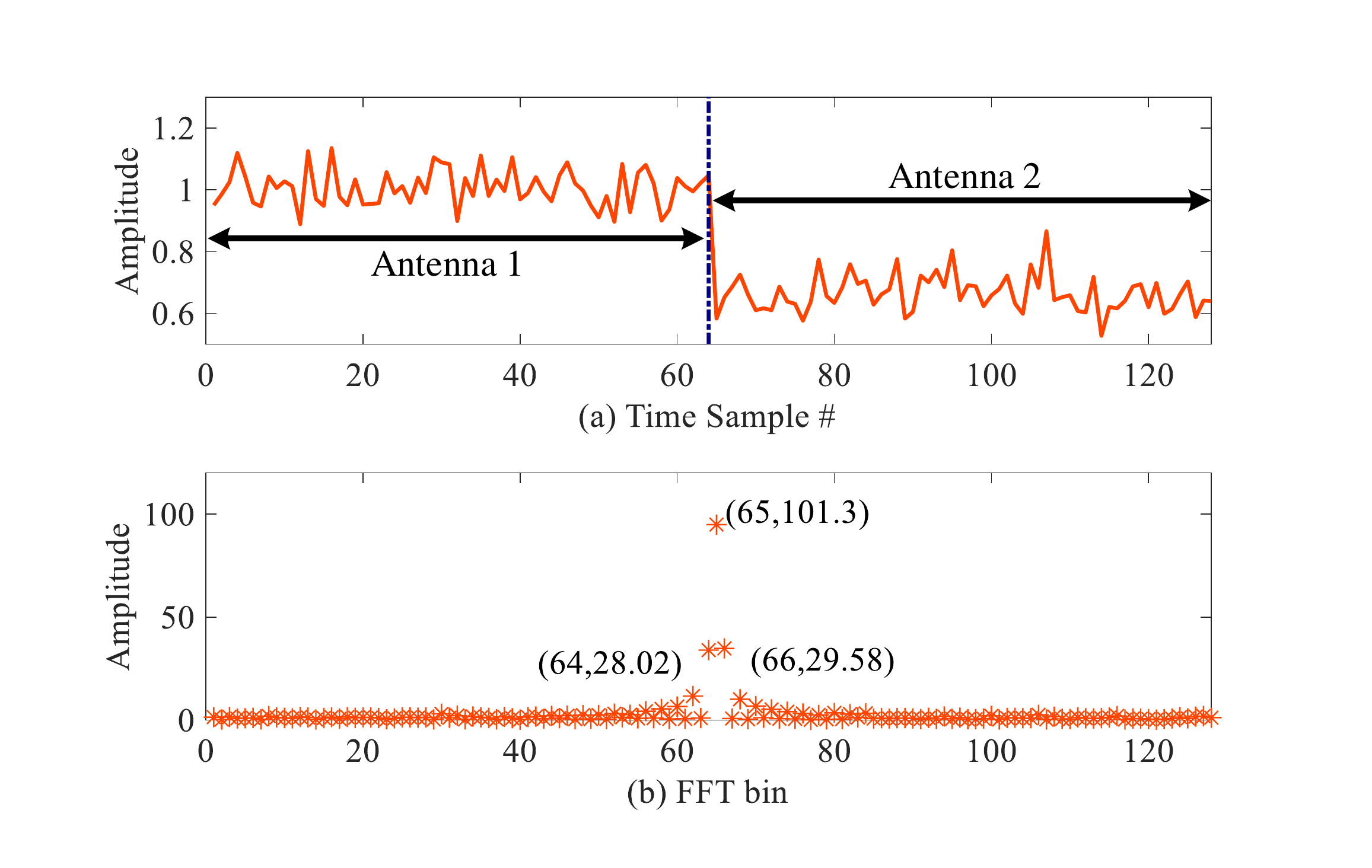} \vspace{-0.3cm}
	\caption{An example of signals' discontinuities. (a) shows the amplitudes of tag's two antennas are different. (b) demonstrates the step signal results in  spectrum leakage.}
	\label{leakage}
	\vspace{-0.35cm}
\end{figure}
So far, we have discussed how to obtain the LLR of received bits in an ideal channel. Next, we consider to computing the LLR in practical channel.

Note that in PLoRa, tags use two antennas to send the bit ``1''. However, due to the channel difference between two antennas, the received signals exhibit discontinuity, which results in spectrum leakage, \textit{i.e.}, the power of  signal splits into peak's adjacent bins. As illustrated in Fig. 7, where the tag sends ``1'', the signal amplitude is discontinuous and the step signal leads to spectrum leakage. To more accurately compute the transition probability, we need to take the leakage into account. 

We observe that the leakage mainly appears in the two FFT bins near the peak. We use the power sum of the $(\overline{s}_i-1)$th, $\overline{s}_i$th and $(\overline{s}_i+1)$th FFT bins to approximate the signal power of bit ``1''. Thus, the LLR of the received bits can be approximated as 

\begin{equation}\label{LLR5}
\begin{aligned}
\begin{split}
L_{i} &\approx \dfrac{|\hat{{f}}_{i,s_i}|^2-\sum_{m=\overline{s}_i-1}^{\overline{s}_i+1}|\hat{{f}}_{i,m}|^2}{2\sigma^2} \ln \dfrac{\sqrt{\sum_{m=\overline{s}_i-1}^{\overline{s}_i+1}|\hat{{f}}_{i,m}|^2}}{|\hat{{f}}_{i,s_i}|}. 
\end{split}
\end{aligned}
\end{equation}

\textbf{LLR combination and information bit decision.}
After estimating the LLR of the received bits, we use the BP algorithm\cite{BP} to compute the LLR of information bits.  Let $L_{q} \left[ t \right]$ denote the LLR for $q$th information bit obtained by the $t$th packet. We combine the LLR of information bit from different packets, and then decode the information bits based on the decision function
\begin{equation}
\hat x_q=\left\{
\begin{array}{lr}
0, if \; \sum_{t=1} ^ T  L_{q} \left[ t \right]   \geq 0\\
1, \text{otherwise},\\
\end{array}\right.
\end{equation}
where the $T$ denotes the number for the receipted packets. With the LLR combination, all redundant bits can be used to decode the information bits.  When no error appear in information bits after looking at
the CRC checksum, the gateway will return an acknowledgment (ACK) frame to the tag to terminate the transmission.

\subsection{Protocol Design}
\begin{figure}[t]
	\centering
	\includegraphics[width=2.7in]{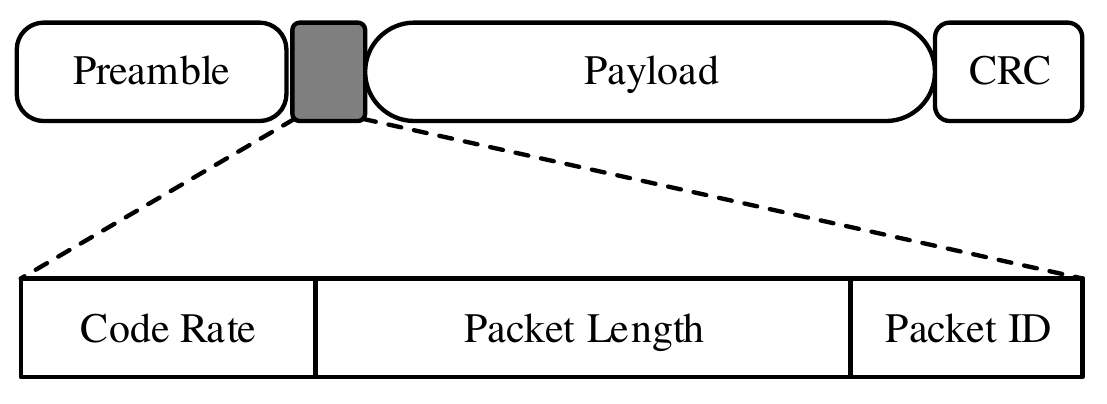} \vspace{-0.1cm}
	\caption{Packet format in PolarScatter. }
	\label{frameformat}
	\vspace{-0.4cm}
\end{figure}

\textbf{Packet format.} The packet format in PolarScatter is shown in Fig.~\ref{frameformat}. We use two bits to indicate code rate and four bits to indicate packet length, and one bit to indicate packet ID. The Packet ID is set to 0 and 1 for the first and the second packets, respectively. The 16-bit CRC is transmitted at the end of the first packet. 

\textbf{Lightweight feedback mechanism.}
In PolarScatter, the gateway needs to feed back the code rate of the second packet to the tag. To build a  simple and effective feedback mechanism, we adopt existing low-power energy detector hardware circuits\cite{Detector} to detect the gateway's feedback. The gateway uses an alternating ON-OFF keying sequence to represent the code rate of the second packet. Such a feedback mechanism only consumes power between 98 nW and 2.4  $\mu$W.

\begin{figure*}
	\centering
	\hspace{-0.85cm}
	\begin{minipage}[t]{0.39\textwidth}\centering
		\centering
		\includegraphics[width=0.9\textwidth]{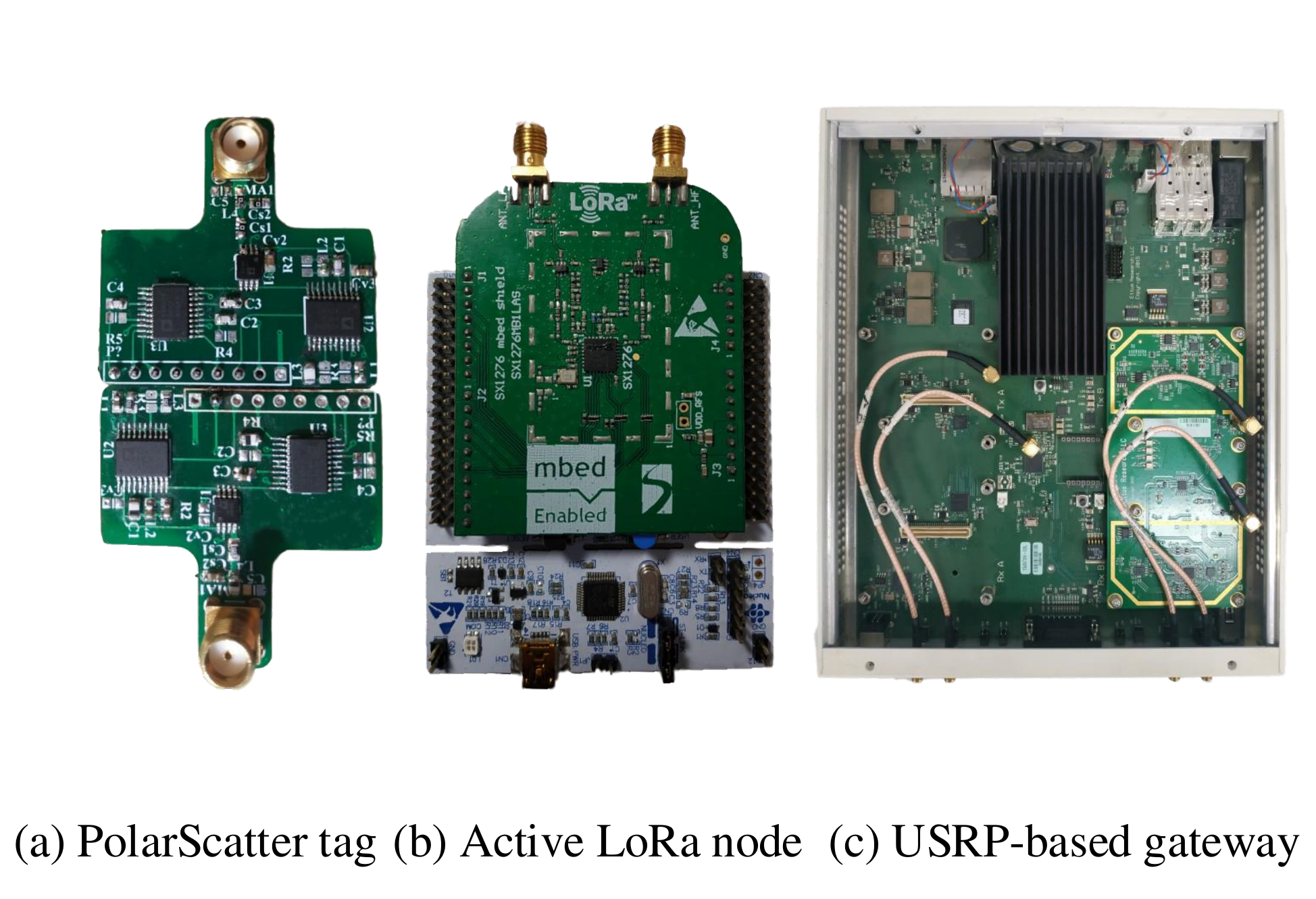} \hspace{-0.8cm}
		\caption{Experimental platform.} 
		\label{Experimenthardware}
	\end{minipage}
	\hspace{-0.3cm}
	\begin{minipage}[t]{0.65\textwidth}\centering
		\centering
		\includegraphics[width=0.9\textwidth]{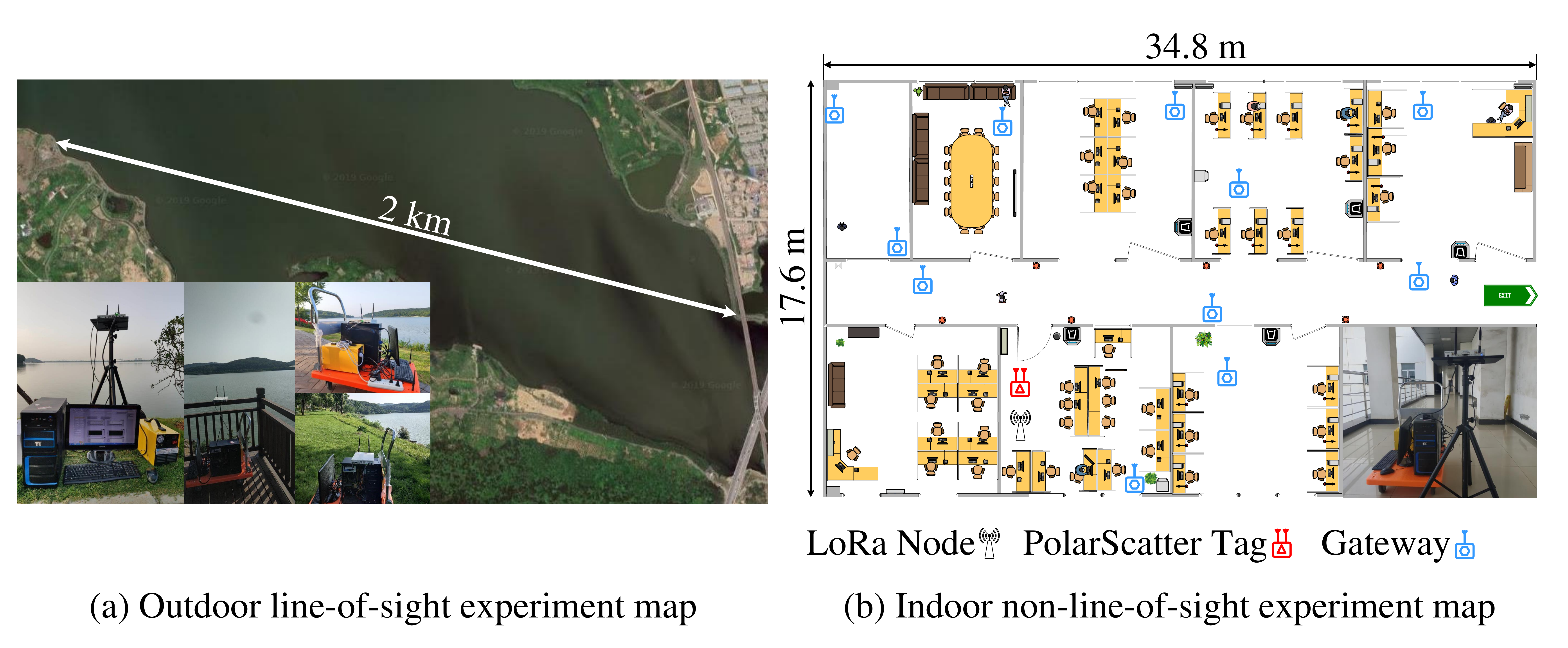}   
		\caption{Indoor and outdoor experimental field.} \label{MAP}
	\end{minipage}
	\vspace{-0.4cm}
\end{figure*}

\section{Implementation} \label{sec:Implementation}

\textbf{PolarScatter tag}.
We design tags according to \cite{peng2018plora} and \cite{tag}.  The tags use two omnidirectional antennas with 3 dBi gain as shown in Fig.~\ref{Experimenthardware}(a). We adopt Microsemi's IGLOO nano AGLN250\cite{IGLOO} as the digital signal processor that runs the polar encoding and baseband modulation components.

\textbf{LoRa node.} We implement LoRa nodes based on SX1276 frontend \cite{SX1276} and Nucleo-L152RE processing unit\cite{Nucleo} as shown in Fig.~\ref{Experimenthardware}(b).
The output power is set to be 20 dBm. The LoRa node works at 900 MHz frequency with 500 kHz modulation bandwidth. The spreading factor (SF)  of LoRa signal can be configured as from 7 to 12. 

\textbf{PolarScatter gateway.}  We prototype a PolarScatter gateway on a USRP X310\cite{USRP} that runs a modified open source LoRa decoder. The USRP is connected to a PC with 8 GB memory and Intel i7 dual-core processor. The USRP adopts an omnidirectional VERT 2450 antenna with 3 dBi gain\cite{VERT2450}. The sample rate of the USRP is configured with 10 MHz to simultaneously monitor six LoRa channels. The decoding algorithm of polar codes is implemented on the PC, and can be easily ported to a more cost-effective gateway.

\section{Evaluation} \label{sec:Evaluation}
In this section, we test PolarScatter's performance in LoS and NLoS scenarios as shown in Fig.~\ref{MAP}(a) and Fig.~\ref{MAP}(b), respectively. We first present the experimental setting and methodology, and then evaluate the effectiveness of \textit{Sozu} polar codes. Next, we show PolarScatter' PRR and goodput.  After that,  we examine the proposed LLR metric. Finally, we demonstrate the storage overhead and power consumption of our proposed polar encoder. 

\textbf{Experimental setup.} 
We test PolarScatter's performance in an outdoor scenario as shown in Fig.~\ref{MAP}(a). The tags, the active LoRa node, and the gateway are deployed without being blocked by any object. We evaluate PolarScatter's performance at 20 locations with different communication distances between the tags and the gateway. The maximum communication distance is 2 km and the
minimum distance is 75 m. The experiments are performed over a week and the weather is mostly clear.

The indoor experimental setup is shown in Fig.~\ref{MAP}(b). We put an active LoRa node and a PolarScatter's tag in an office and move the gateway through a corridor and several adjacent meeting rooms and offices. The PolarScatter tag is marked as a red dot, and the locations of the gateway are marked as blue dots.  For each location, we collect about 6000 packets. 

\textbf{Methodology.} We compare the performance of PolarScatter with PLoRa, which is the state-of-the-art long-range backscatter. For the sake of fairness, we add Hamming coding to PLoRa. The Hamming code encodes four bits of data into seven bits by adding three redundant bits. We collect various tags' data and compare PolarScatter and PLoRa using two metrics: PRR and goodput. 

\subsection{Effectiveness of \textit{Sozu} Polar Codes}
\begin{figure}[t]
	\vspace{0.15cm}
	\centering
	\includegraphics[width=2.7in]{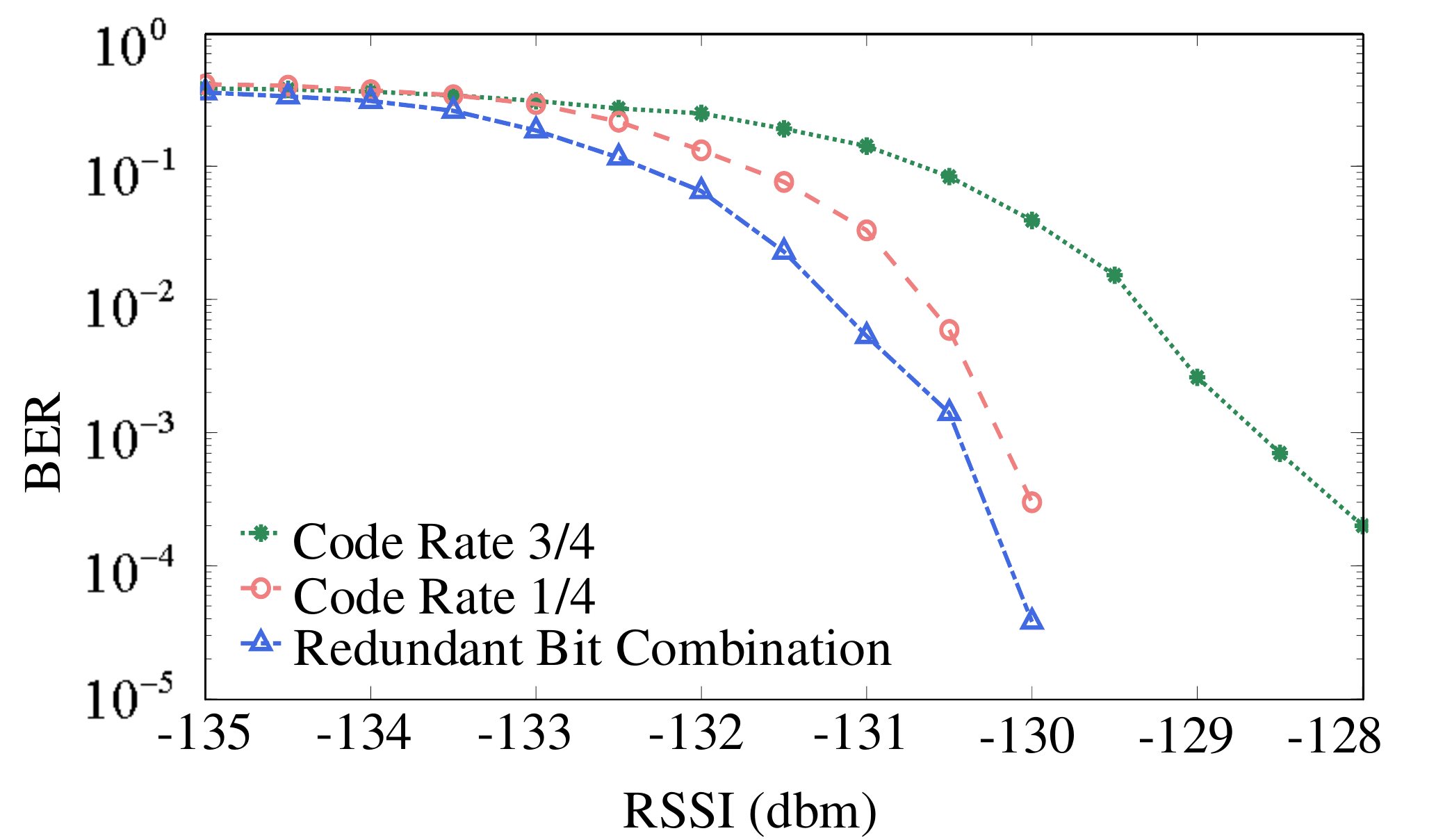} \vspace{-0.1cm}
	\caption{Redundant bit combination. This figure shows we can boost the decoding performance by combining redundant bits from different packets.} 
	\label{LLRcombination}
	\vspace{-0.3cm}
\end{figure}

We first examine  effectiveness of \textit{Sozu} polar codes.
In this experiment, we keep an active node and a tag stationary, while moving the gateway to different locations in order to obtain different received signal strength indicators (RSSIs). In each transmission, we first transmit the information bits, and then add redundant bits with 3/4 and 1/4 code rates, respectively.  We then combine redundant bits of two code rates to perform polar decoding. We report the bit error rate (BER) of different schemes. The test results are shown in Fig.~\ref{LLRcombination}. We can see that the redundant bit combination scheme can effectively reduce the BER. This means that \textit{Sozu} polar codes can improve the channel reliability by combining redundant bits from different packets.

\subsection{PRR and Goodput}
We evaluate PRR and goodput of PolarScatter in both LoS and NLoS scenarios.  The goodput of a received frame with the bit rate $R$ can be calculated as
\begin{equation}\label{goodput}
\begin{split}
Goodput = \dfrac{N_{cleanbit}}{N_{totalbit}} \times R,
\end{split}
\end{equation} 
where $N_{cleanbit}$ and $N_{totalbit}$ represent the number of bits that correctly received packets contain and the number of total bits, respectively.

\textbf{Outdoor scenario.}
In this experiment, we keep the source (the active LoRa dode) and the tag stationary. The distance between the source and the tag is 1 m. We move the gateway to different locations.  We interleave PolarScatter and PLoRa to ensure that the two systems experience similar channels quality. 
We then test the PRR and goodput of the two systems, respectively. 

Fig.~\ref{PRR_LOS} presents the PRR under different communication
distances with SF = 7 and SF = 12. The maximum communication distance of all other SFs are between these two curves. Compared with PLoRa, PolarScatter significantly improves the PRR. With SF = 7 and PRR = 0.5, PolarScatter can extend the communication range from 80 m to 300 m. Similarly, with SF = 12 and PRR = 0.5,  PolarScatter can extend the communication range from 350 m to 1500 m. 

Fig.~\ref{Throughput_LOS} shows the goodput under different communication ranges.  It can be seen that PolarScatter significantly boosts the goodput. Compared with PLoRa, PolarScatter achieves 2$\times$ to 10$\times$ goodput gain.  
In the SF = 7 case, when the communication distance is beyond 225 m, the PLoRa's goodput is close to 0, while PolarScatter still provides several hundreds~bps goodput. In the SF = 12 case, the maximum transmission distances of PLoRa and PolarScatter are about 1.1 km and 2 km, which shows PolarScatter can effectively extend the communication limit. 

\begin{figure}
	\centering
	\includegraphics[width=2.7in]{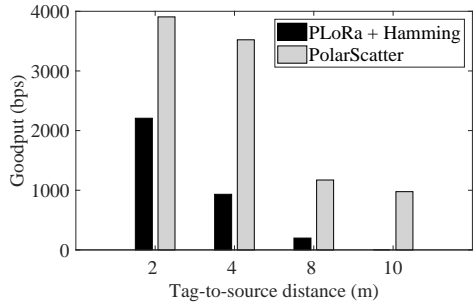} \vspace{-0.2cm}
	\caption{Goodput versus distance between source and tag. This figure shows significantly improve the goodput of long-range backscatter} 
	\label{goodputVsdistance}\vspace{-0.2cm}
\end{figure}

\textbf{Indoor scenario.} 
We repeat the previous experiments in an indoor office building. The distance between the source and the tag is 3 m. SF and BW are set to be 7 and 500~kHz, respectively. The test results are summarized in Fig.~\ref{NLOSresults}. The results reveal that PolarScatter significantly outperforms PLoRa in goodput and PRR.
As shown in Fig.~\ref{NLOSresults},  the PLoRa's goodput of the nodes 6, 7, 8, 9 and 10 are close to 0. However, PolarScatter still can communicate at several hundred bps. 

The above results demonstrate that PolarScatter can make effective utilization of backscatter link capacity, and thus achieves high-speed connectivity when communicating over long distances.

\subsection{Goodput versus Tag-to-Source Distance}

In this experiment, we place one active LoRa node as the excitation signal generator (called source). 
The  source and gateway are kept stationary and separated with a distance of 100 m. We move a PolarScatter tag to different locations, and then report the goodput. The test results are shown in Fig.~\ref{goodputVsdistance}.   When the tag is placed near the source (\textit{e.g.}, 2 m away), the PolarScatter and PLoRa maintain similar throughput. However, PolarScatter significantly outperforms PLoRa in tag-to-source distances of more than 4 m.  This result indicates that PolarScatter can cope with very challenging channel conditions and significantly improve the goodput of long-range backscatter.

\subsection{ LLR of Received Bits}
We examine the effectiveness of our proposed method in calculating LLR of received bits by leveraging MATLAB simulation.  We compare our proposed method with the conventional method that computes the LLR based on the signal power. For each experiments, we collect 1000 packets and report the cumulative distribution function (CDF) of LLR under SNR = -15 dB and SNR = -20 dB, respectively.

Fig.~\ref{CTP} shows the results.  It is clear that our proposed method is more close to the theoretical distribution, compared with the conventional method. We  can also see that the smaller
the SNR is, the more inaccurate the conventional method is. This is because it is very hard to obtain accurate signal power at low SNR. In contrast, since our method is not affected by the signal power, it works well in low SNR. 

\begin{figure}[t]
	\begin{minipage}[t]{0\linewidth}
		\centering
		\subfigure[SF=7]{	
			\label{PRR_SF7}	
			\includegraphics[width=1.85in]{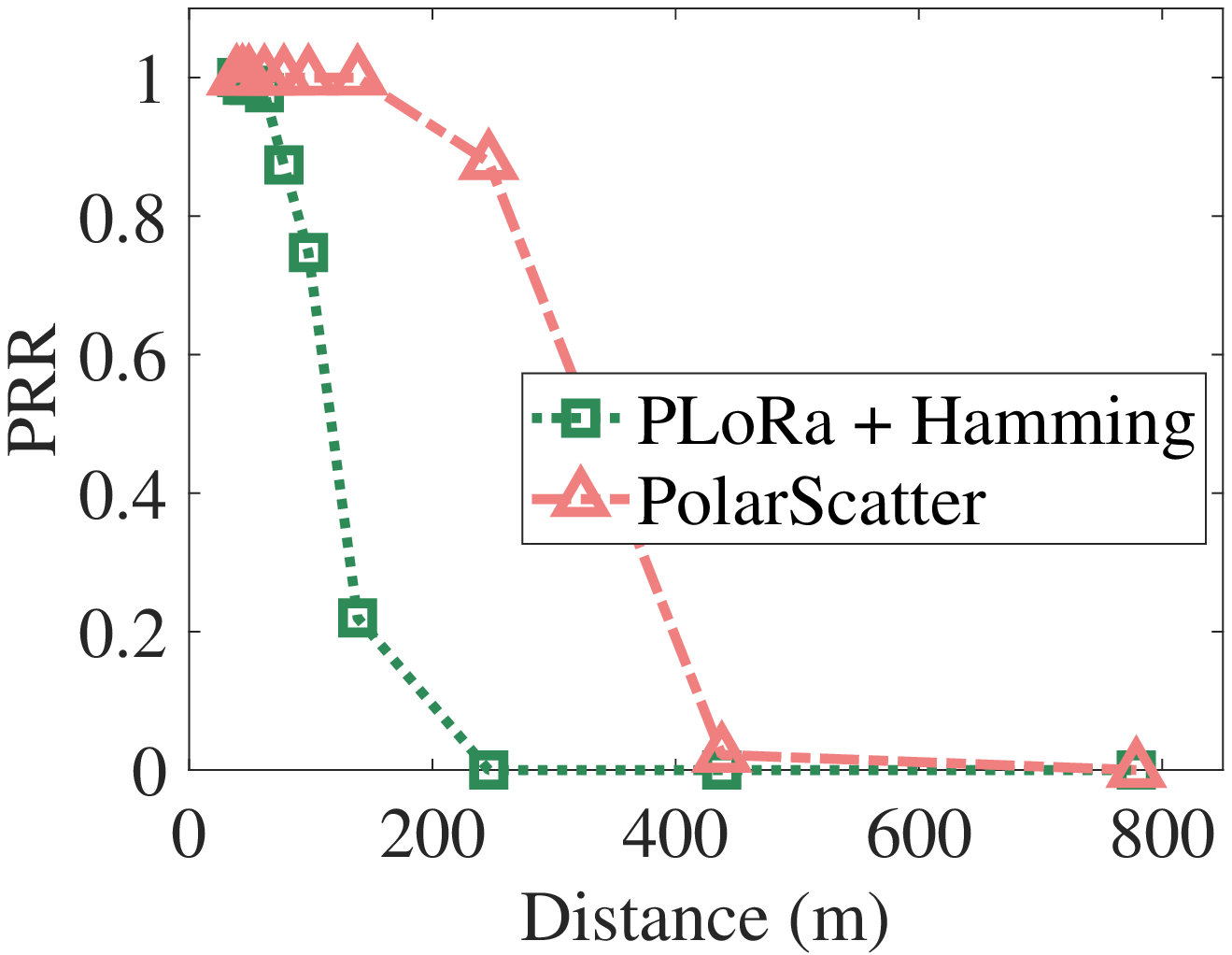}}
	\end{minipage}
		\hspace{-0.8in} 
	\begin{minipage}[t]{2\linewidth}
		\centering
		\subfigure[SF=12]{	
			\label{PRR_SF12}	
			\includegraphics[width=1.85in]{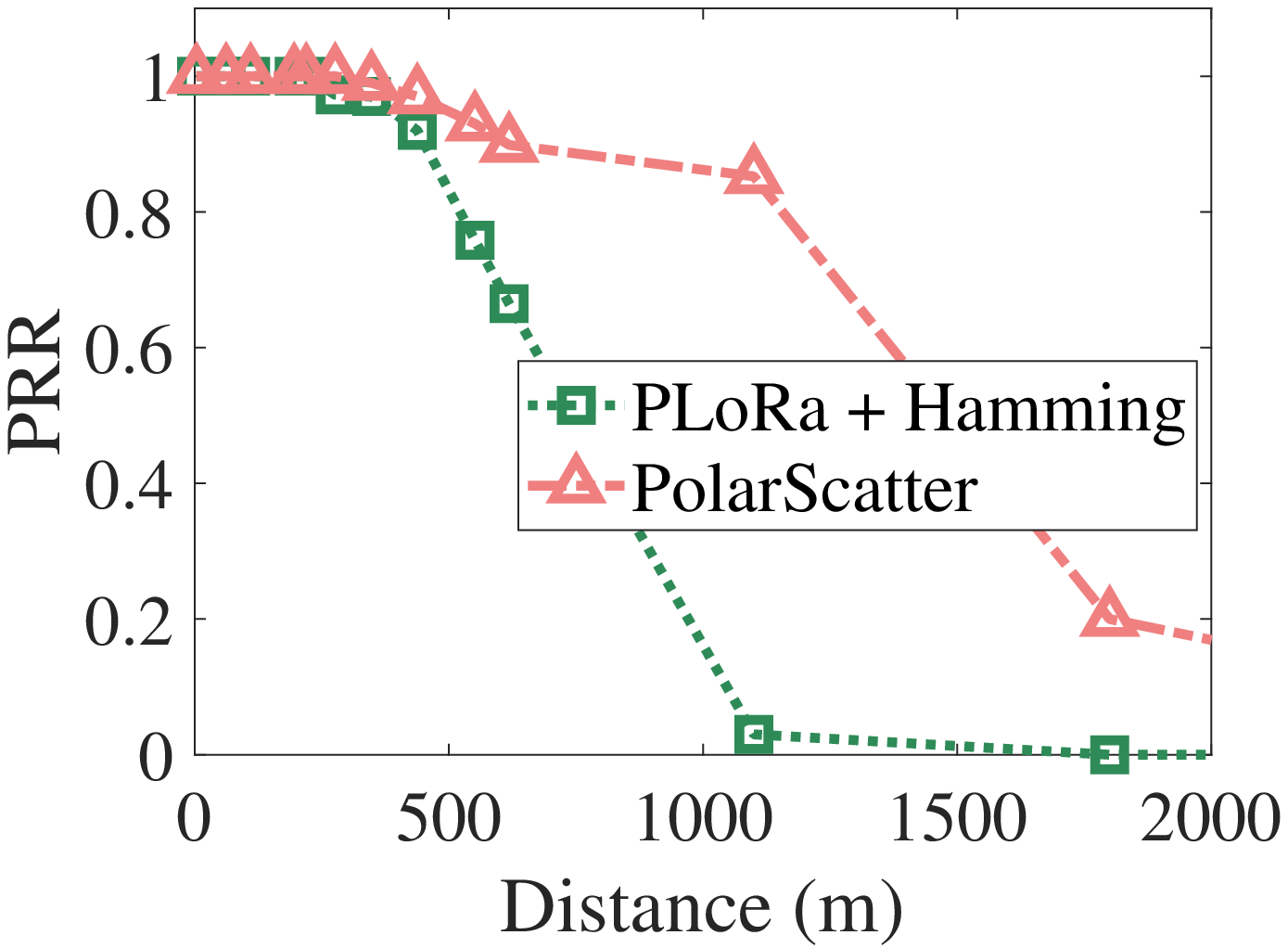}}
	\end{minipage}
	\caption{Packet reception rates under different communication distances. It is clear that the PolarScatter can significantly boost PRR.}
	\label{PRR_LOS}
	\vspace{0cm}
\end{figure}

\begin{figure}[t]
	\begin{minipage}[t]{0\linewidth}
		\centering
		\subfigure[SF=7]{	
			\label{ThroughPut_SF7}	
			\includegraphics[width=1.85in]{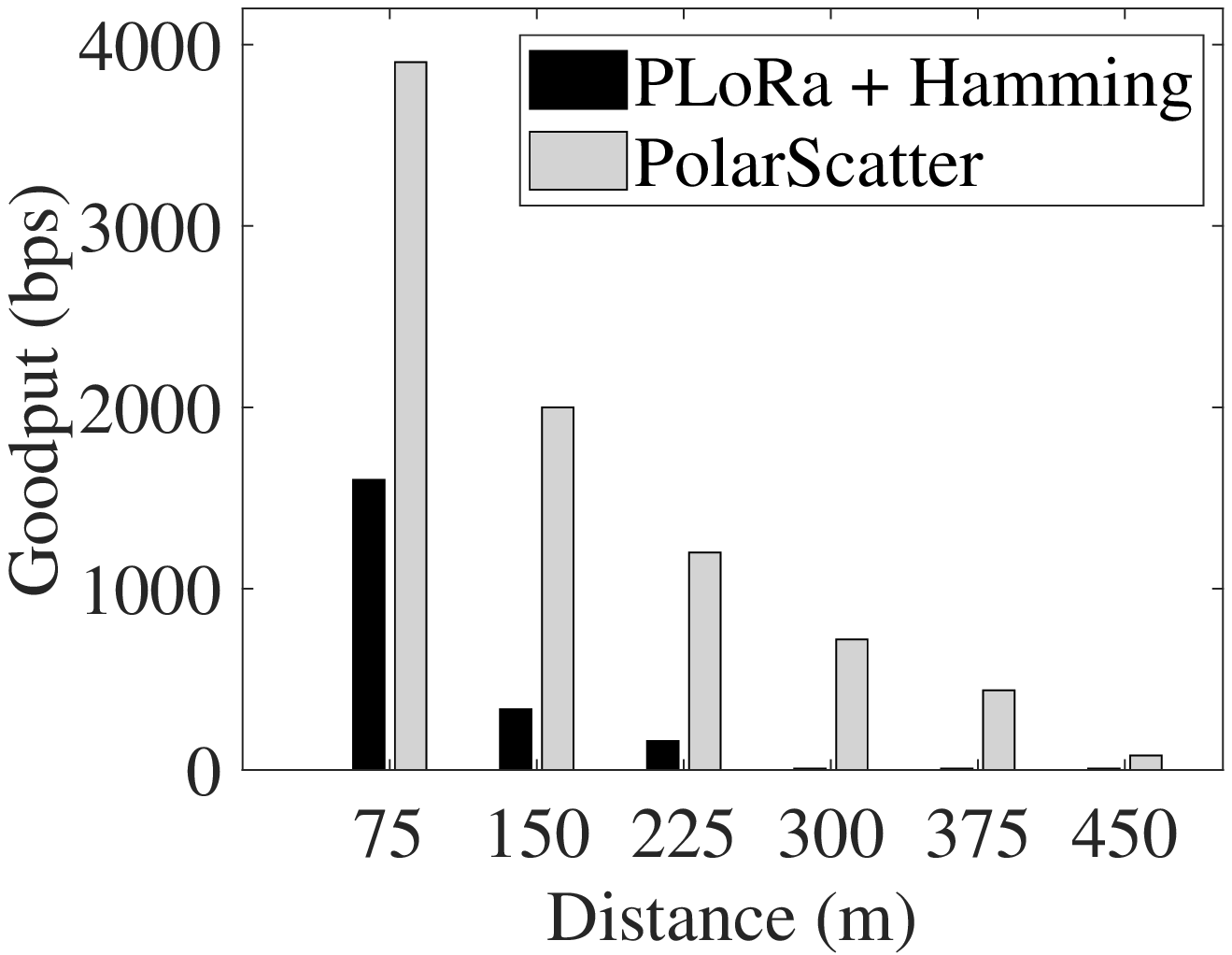}}
	\end{minipage}
	\hspace{-0.8in}
	\begin{minipage}[t]{2\linewidth}
 
		\centering
		\subfigure[SF=12]{	
			\label{ThroughPut_SF12}	
			\includegraphics[width=1.85in]{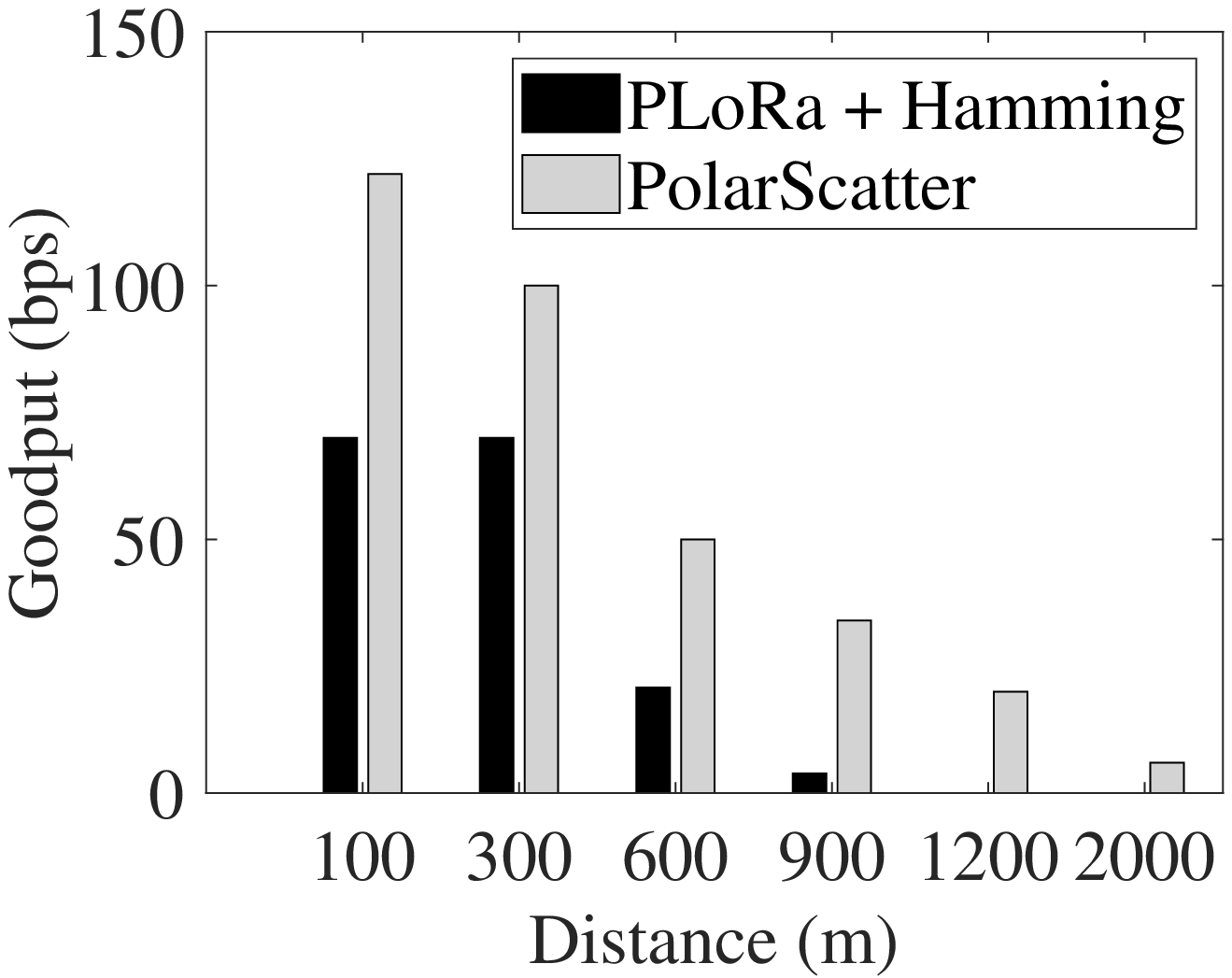}}
	\end{minipage}
	\caption{Goodput under different communication distances. This figure shows that PolarScatter can significantly improve goodput.}
	\label{Throughput_LOS}
	\vspace{-0.3cm}
\end{figure}

\begin{figure}[t]
	\begin{minipage}[t]{0\linewidth}
		\centering
		\subfigure[PRR]{	
			\label{ThroughputNLOS}	
			\includegraphics[width=1.8in]{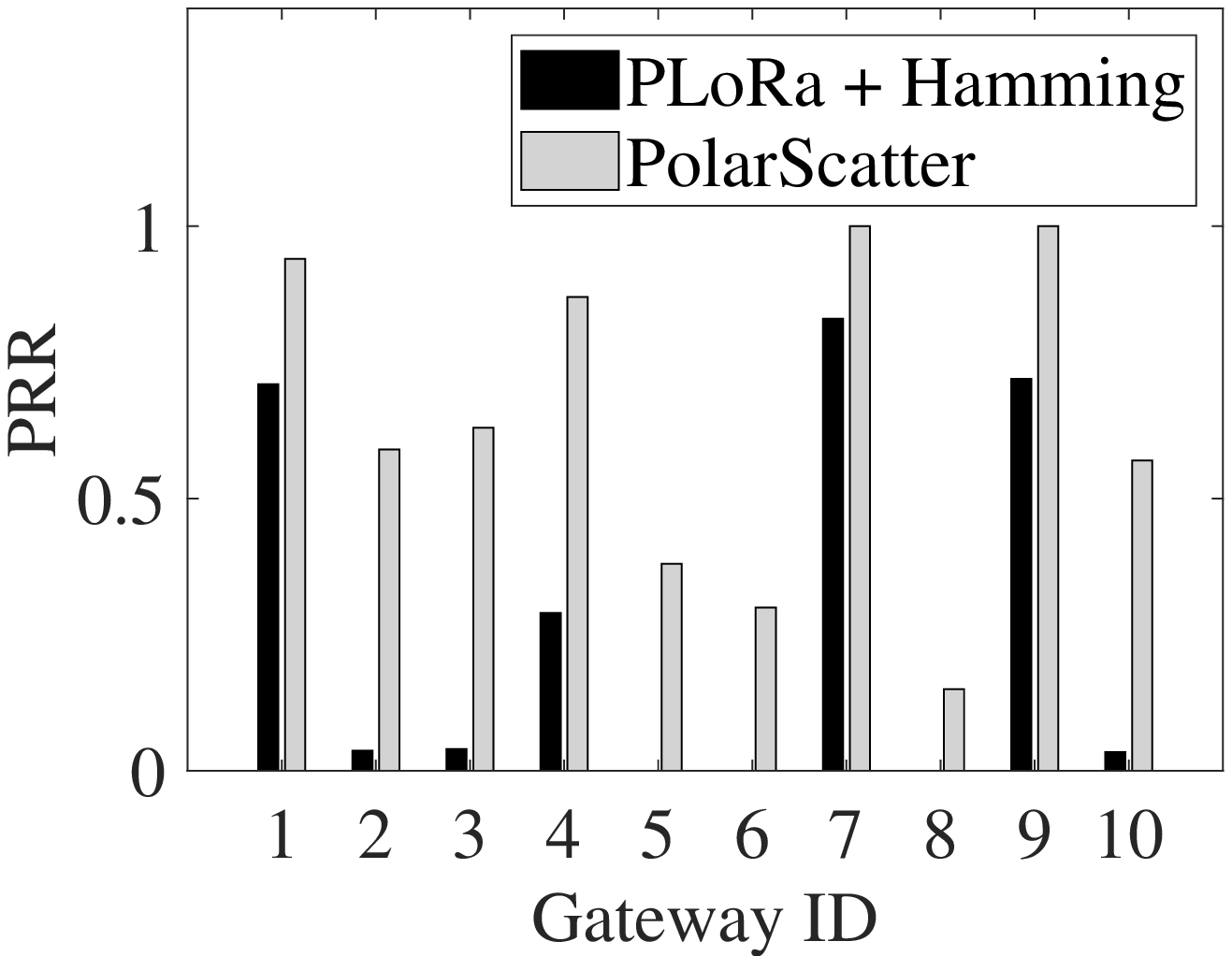}}
	\end{minipage}
		\hspace{-0.8in} 
	\begin{minipage}[t]{2\linewidth}
		\centering
		\subfigure[Goodput]{	
			\label{PRRNLOS}	
			\includegraphics[width=1.8in]{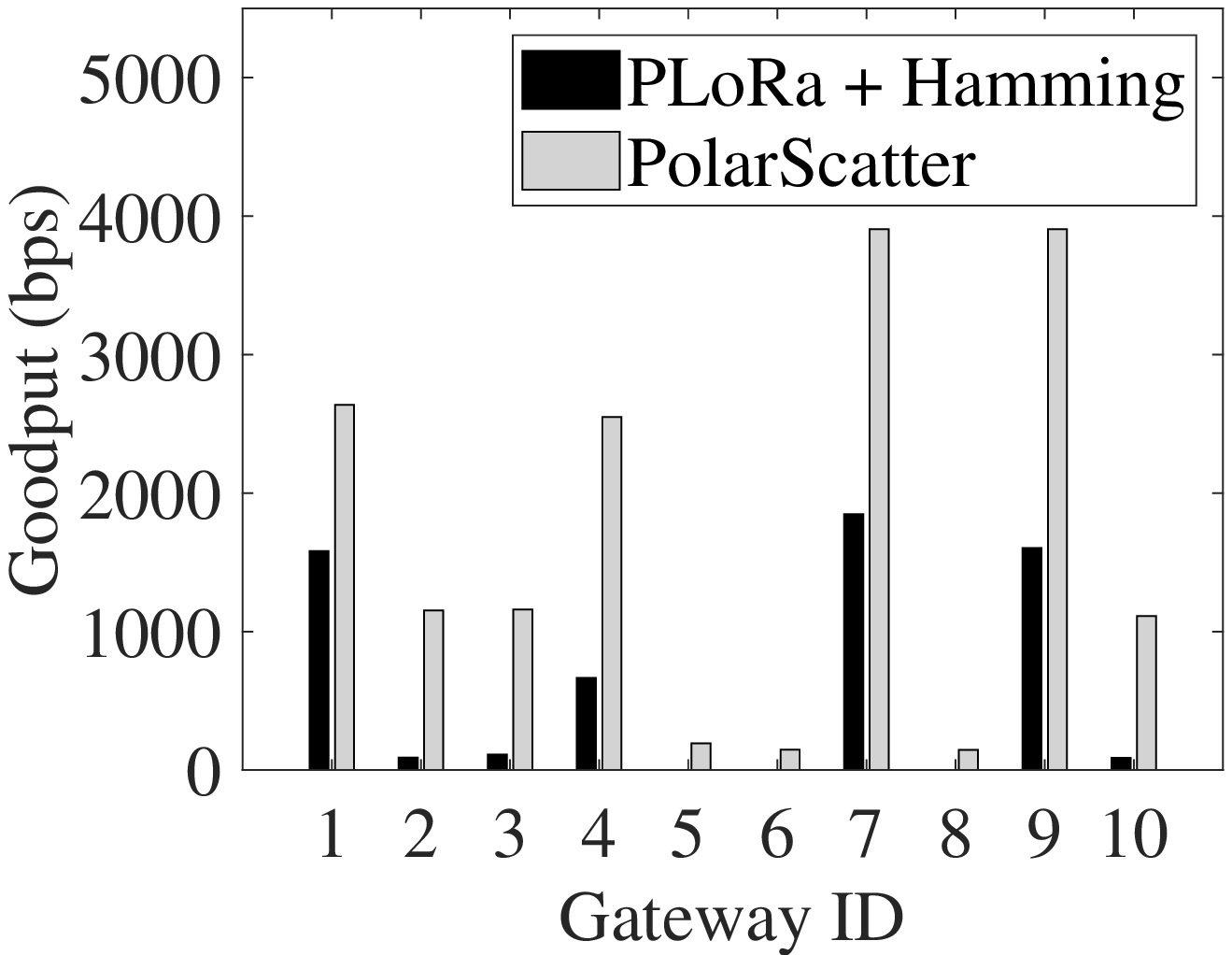}}
	\end{minipage}
\vspace{-0.1cm}
	\caption{Decoding performance comparison at different node locations in an indoor scenario. This figure show that PolarScatter outperforms PLoRa in PRR and throughput}
	\label{NLOSresults}
	\vspace{-0.3cm}
\end{figure}

\subsection{Energy Consumption and Storage Overhead}
We simulate an IC design of polar encoder in TSMC 65 nm LP CMOS process \cite{IC} based on industry-standard EDA tools\cite{Cadence,Synopsis}. We implement our polar encoder and the conventional polar encoder with four code lengths, \textit{i.e.}, 128, 256, 512, and 1024, respectively. We compare the two encoders using two metrics: storage overhead and power consumption.

\begin{figure}[t]
	\begin{minipage}[t]{0\linewidth}
		\centering
		\subfigure[SNR = -15 dB]{	
			\label{CTP15db}	
			\includegraphics[width=1.8in]{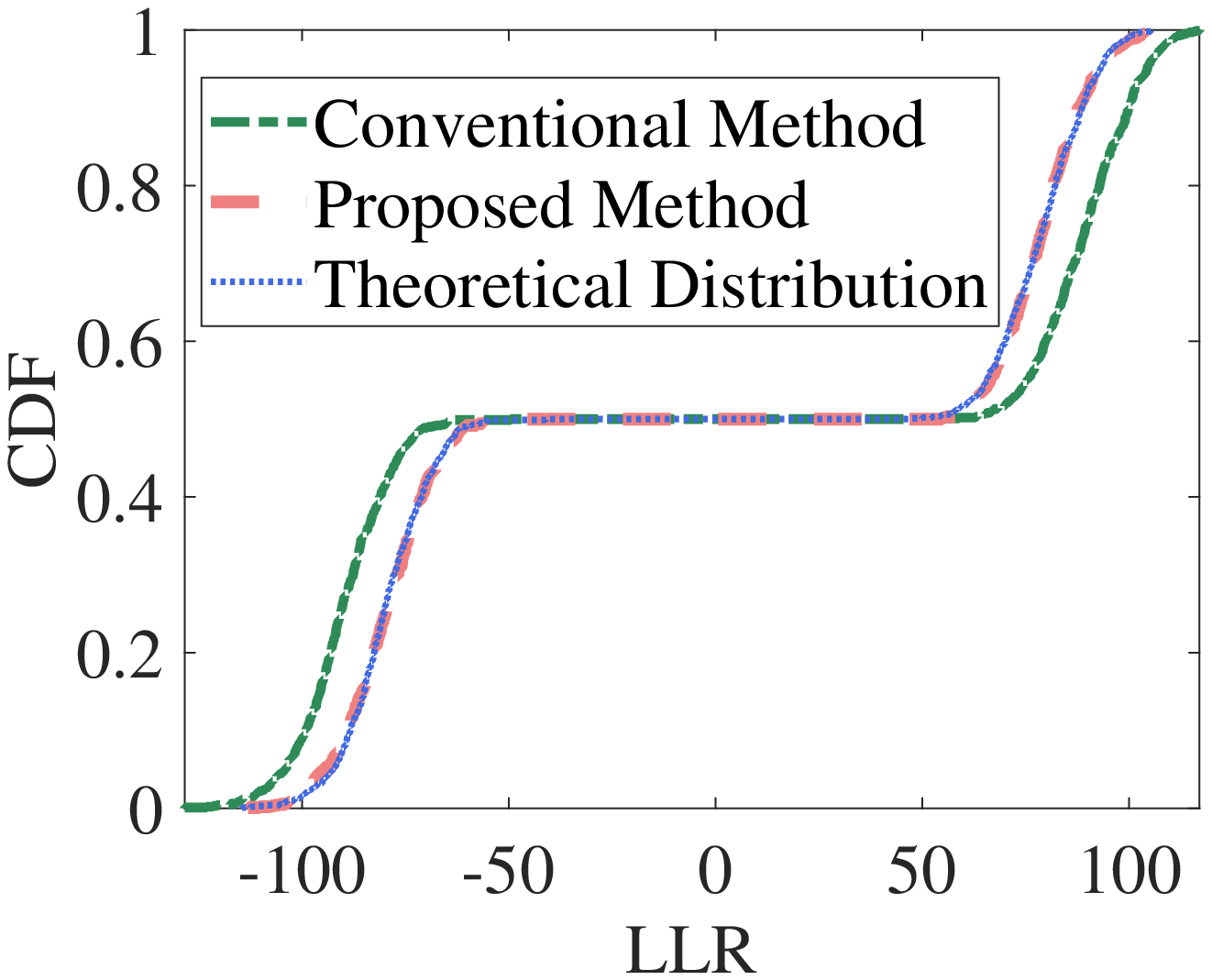}}
	\end{minipage}
		\hspace{-0.8in} 
	\begin{minipage}[t]{2\linewidth}
		\centering
		\subfigure[SNR = -20 dB]{	
			\label{CTP30db}	
			\includegraphics[width=1.8in]{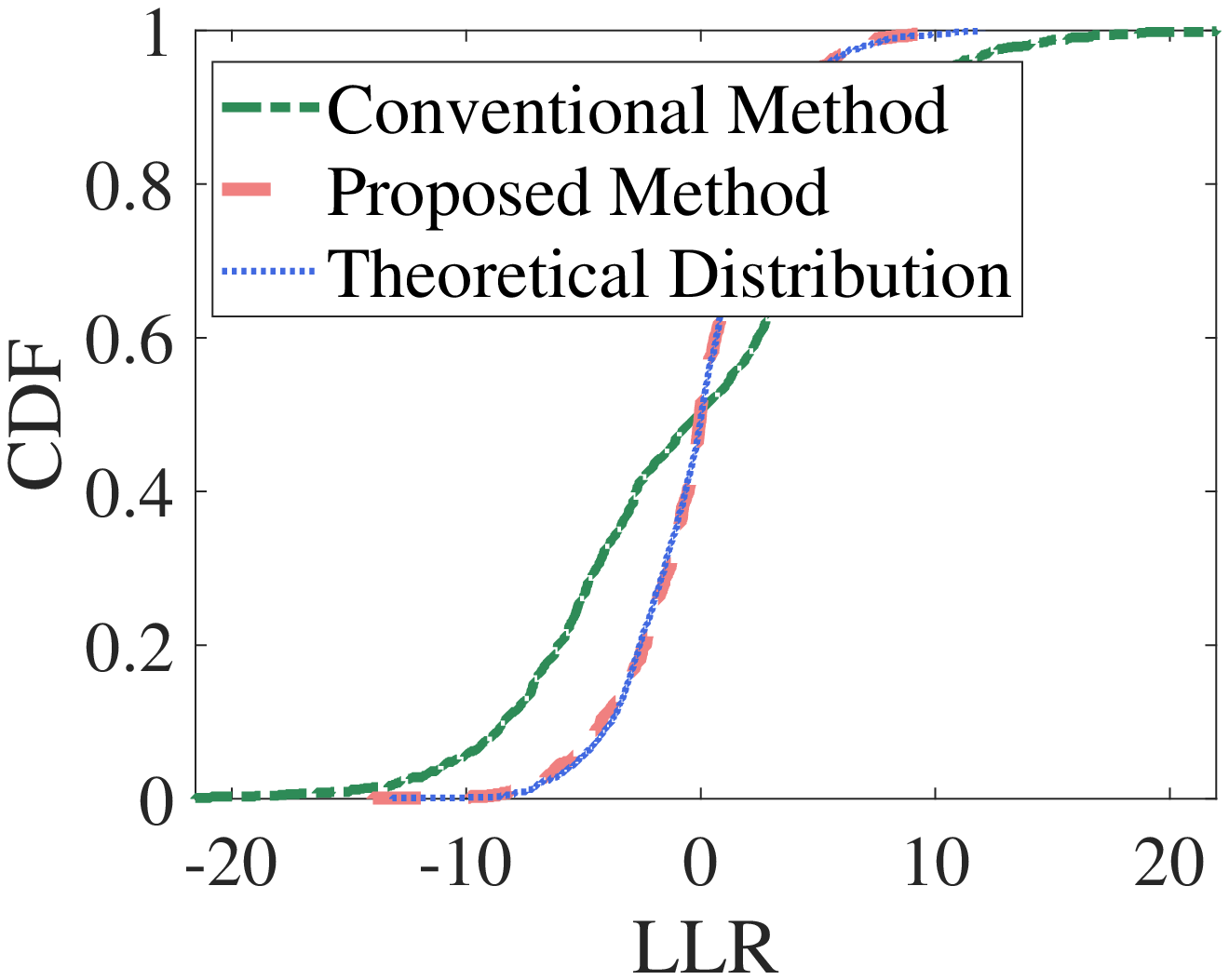}}
	\end{minipage}
	\caption{CDF of LLR with different algorithms. This figure shows that our proposed method outperforms the conventional method.}
	\label{CTP}
	\vspace{-0.2cm}
\end{figure}

We first evaluate the storage overhead and the results are shown in Table \ref{table1}. It can be seen that our encoder can significantly reduce the storage overhead.
Compared with the conventional encoder, our encoder can decrease the storage overhead from 35$\times$ to 465$\times$. 
For code length $N=1024$, our encoder can lower the storage overheads from 2.3 MB to 5KB. 

We proceed to evaluate the power consumption of our encoder. The results in Table \ref{table2} show our polar encoder can significantly reduce power consumption. For code length $N=1024$, the conventional encoder consumes 1870 $\mu$W of power, while our encoder can work at tens of microwatts of power. With such low power consumption, our encoder can be easily applied to batteryless backscatter tags. 

\begin{table}[htbp]
	\centering  
	\caption{Storage overhead for different code lengths}  
	\label{table1}  
	\begin{tabular}{lllll}
		\hline
		\multicolumn{1}{|l|}{Code lengths}    & \multicolumn{1}{c|}{128}& \multicolumn{1}{l|}{256}& \multicolumn{1}{l|}{512}& \multicolumn{1}{l|}{1024}\\ \hline
		\multicolumn{1}{|l|}{Conventional encoder} & \multicolumn{1}{l|}{18 KB}& \multicolumn{1}{l|}{144 KB}& \multicolumn{1}{l|}{576 KB}& \multicolumn{1}{l|}{2322 KB}\\ \hline
		\multicolumn{1}{|l|}{Our encoder} & \multicolumn{1}{l|}{0.5 KB} & \multicolumn{1}{l|}{1 KB} & \multicolumn{1}{l|}{2 KB} & \multicolumn{1}{l|}{5 KB} \\ \hline
		&&&&                                   
	\end{tabular}
	\vspace{-0.6cm}
\end{table}

\begin{table}[htbp]
	\centering  
	\caption{Power consumption for different code lengths}  
	\label{table2} 
	\setlength{\tabcolsep}{2mm}{
		\begin{tabular}{lllll}
			\hline
			\multicolumn{1}{|l|}{Code lengths} & \multicolumn{1}{c|}{128}& \multicolumn{1}{l|}{256} & \multicolumn{1}{l|}{512}& \multicolumn{1}{l|}{1024}\\ \hline
			\multicolumn{1}{|l|}{Conventional encoder} & \multicolumn{1}{l|}{67 $\mu$W}& \multicolumn{1}{l|}{372 $\mu$W}& \multicolumn{1}{l|}{811 $\mu$W}& \multicolumn{1}{l|}{1780 $\mu$W}\\ \hline
			\multicolumn{1}{|l|}{Our encoder} & \multicolumn{1}{l|}{11 $\mu$W} & \multicolumn{1}{l|}{20 $\mu$W} & \multicolumn{1}{l|}{37 $\mu$W} & \multicolumn{1}{l|}{71 $\mu$W} \\ \hline
			&&&&                                   
	\end{tabular}}
	\vspace{-0.5cm}
\end{table}

\section{Related Work} \label{sec:relatedwork}

\textbf{Long-range Backscatter.}
In recent years, some pioneering systems have been proposed to support long-range backscatter communications\cite{peng2018plora,talla2017lora,varshney2017lorea}. LoRa backscatter\cite{talla2017lora}  achieves long-range backscatter communications by generating LoRa signals at the backscatter tags. PLoRa\cite{peng2018plora}  takes ambient LoRa transmissions as excitation signals. LoRea\cite{varshney2017lorea} backscatters narrow-band single tone signals to deliver backscatter signals over long ranges. These innovations focus on hardware design and signal modulation. In contrast, PolarScatter is a physical layer coding design and aims to make effective utilization of backscatter link capacity, thereby providing high throughput when communicating over long ranges. PolarScatter can be easily applied to existing backscatter designs without hardware modifications. 

\textbf{Rate Adaptation in Backscatter.}
There are extensive efforts in improving backscatter throughput by selecting a proper bit rate for different channel quality\cite{RAB,CARA,Blink}. 
RAB \cite{RAB}, CARA\cite{CARA} and  Blink \cite{Blink} require the receiver to estimate RSSI. These technologies work pretty well in short-range backscatter systems. However, in long-range backscatter, it is difficult to obtain accurate RSSI,
since the received signal is drowned by noise.
PolarScatter leverages polar codes to make effective use of correct bits to boost throughput when communicating over long distances.  Instead of relying on RSSI, PolarScatter automatically adjusts to an appropriate code rate based on FBER, which works well in extremely low SNR conditions.

\textbf{Polar Codes.}
In the past decade, there has been a large body of theoretical work that boosts decoding performance by polar codes, and shows it can approach Shannon limit\cite{arikan2009channel,HARQ1,HARQ2,HARQIoT}.
Zhao \textit{et al.}\cite{HARQ1}  present rate compatible polar codes through polarizing matrix extension. Saber \textit{et al.}\cite{HARQ2} propose a novel puncturing algorithm to extend polar codes.  Mohammadi \textit{et al.}\cite{HARQIoT} leverage systematic polar codes to adapt the rate of transmission in IoT applications. 
While these polar code schemes differ in detail, all are designed for active radios, and are unaffordable to low-end backscatter tags.  Unlike traditional polar codes, the proposed \textit{Sozu} polar codes do not require accurate channel. Furthermore, it reduces storage overhead of the coding matrix and energy consumption by orders of magnitudes, making it feasible on batteryless backscatter tags.

\section{Conclusion}  \label{sec:Conclusion}
This paper presents PolarScatter, a system that enables reliable long-range transmissions for batteryless backscatter by leveraging channel polarization.  We design \textit{Sozu} polar codes to best exploit the link capacity and automatically adjust to a suitable effective bit rate for different channel quality. 
Furthermore, we propose a low-cost encoder to accommodate polar codes on resource-constrained tags, and design an LLR  metric to accurately perform polar decoding. 
Experiment results show PolarScatter achieves up to 10$\times$ throughput gain, or extends the range limit by 1.8$\times$ compared with the best known long-range backscatter solution. We also design a low-cost polar encoder IC that reduces storage overhead by three orders of magnitude compared with conventional encoders, and lowers the power consumption to merely tens of microwatts.
While our current implementation of PolarScatter is based on PLoRa, we believe the framework can be
extended to support other types of backscatter technologies to significantly boost throughput and  push the communication range limits.

\balance
\bibliographystyle{IEEEtran}

\end{document}